\documentclass[%
reprint,
 amsmath,amssymb,
 aps,
floatfix
]{revtex4-2}

\usepackage{graphicx}
\usepackage{dcolumn}
\usepackage{bm}
\usepackage{color}

\begin{document}

\markboth{Y.\ Hosaka and S.\ Komura}
{Nonequilibrium Transport Induced by Biological Nanomachines}

\title{Nonequilibrium Transport Induced by Biological Nanomachines
}

\author{Yuto Hosaka}\email{yuto.hosaka@ds.mpg.de}

\affiliation{Max Planck Institute for Dynamics and Self-Organization (MPI DS), Am Fassberg 17, 37077 G\"{o}ttingen, Germany}

\author{Shigeyuki Komura}\email{komura@wiucas.ac.cn}

\affiliation{Wenzhou Institute, University of Chinese Academy of Sciences, Wenzhou, Zhejiang 325001, China}
\affiliation{Oujiang Laboratory, Wenzhou, Zhejiang 325000, China}
\affiliation{Department of Chemistry, Graduate School of Science, Tokyo Metropolitan University, Tokyo 192-0397, Japan}

\begin{abstract}
Biological nanomachines are nanometer-size macromolecular complexes that catalyze chemical reactions in the presence of substrate molecules. The catalytic functions carried out by such nanomachines in the cytoplasm, and biological membranes are essential for cellular metabolism and homeostasis. During catalytic reactions, enzymes undergo conformational changes induced by substrate binding and product release. In recent years, these conformational dynamics have been considered to account for the nonequilibrium transport phenomena such as diffusion enhancement, chemotaxis, and substantial change in rheological properties, which are observed in biological systems.
In this review article, we shall give an overview of the recent theoretical and experimental investigations that deal with nonequilibrium transport phenomena induced by biological nanomachines such as enzymes or proteins.
\end{abstract}

\maketitle

\section{Introduction}
\label{sec:intro0}

In recent years, to better understand the complex nonequilibrium phenomena observed in living systems such as the cytoplasm and biological membranes, various studies have been performed in an interdisciplinary field involving physics, chemistry, biology, and engineering.
Experimental studies demonstrate that by harnessing chemical energy, biological nanomachines give rise to nonequilibrium transport phenomena such as diffusion enhancement, chemotaxis or antichemotaxis, and substantial change in rheological properties.
From a theoretical viewpoint, however, equilibrium concepts such as the time-reversal symmetry, Newton's third law, and the reciprocal relation do not hold in nonequilibrium living systems.

At the nanometer scale, catalytic enzymes change their shapes periodically by converting chemical energy into mechanical work.
Continuous energy consumption breaks the time-reversal symmetry at microscopic scales.
Cyclic shape changes of enzymes induce a fluid flow that can be represented by force dipole or torque dipole in a continuum description.
When multiple enzymes are present, their induced collective hydrodynamic flow can give rise to enhancement in tracer or enzyme diffusion coefficients.
Moreover, enzymes interact with chemical species and it is known that the resultant interaction becomes nonreciprocal, i.e., absence of the action-reaction symmetry.
These hydrodynamic and nonreciprocal interactions of enzymes would account for nonequilibrium transport such as diffusion enhancement, chemotaxis, or active phase separation.

At the macroscopic scale, on the other hand, the system with broken time-reversal symmetry can be viewed as an active fluid where the surrounding environment actively interacts with enzymes or proteins.
In living systems, parity symmetry can also be violated due to chiral structures or rotational motions of enzymatic molecules.
An example of such a system is a biological membrane with embedded rotary proteins.
This active chiral fluid possesses unique properties such as edge flows or nonreciprocal interaction that are also essential for nonequilibrium transport in living systems.

The aim of this article is to review the recent developments in the study of nonequilibrium transport phenomena induced by biological nanomachines. 
In Sec.~\ref{sec:intro}, we explain the fundamental physical properties of biological nanomachines.
In Sec.~\ref{sec:nonequilibrium}, we review recent experimental works dealing with nonequilibrium transport in living systems such as cells. 
In Sec.~\ref{sec:coarse-graind}, we describe several coarse-grained descriptions of biological nanomachines.  
In Sec.~\ref{sec:odd}, we review the recent developments on the concept of odd viscosity that breaks both time-reversal symmetry and parity symmetry. 
Finally, some future perspectives are given in Sec.~\ref{sec:perspectives}.

\section{Physical Chemistry of Biological Nanomachines}
\label{sec:intro}

\subsection{Biological nanomachines}

Biological nanomachines are nanometer-size proteins that catalyze chemical reactions in the presence of substrate molecules, e.g., adenosine triphosphate (ATP)~\cite{albertsbook}.
During chemical reactions, nanomachines or motor proteins change their shapes to exert forces on surrounding environments such as the cytoplasm or biological membranes.
For example, the myosin proteins are responsible for muscle contraction by generating directional movement along actin filaments, while kinesins and dyneins that walk along microtubules are essential for vesicle trafficking.
In addition to these translational motors responsible for motile processes in a biological cell~\cite{phillips2012}, there are rotor proteins called ATP synthase or F$_0$F$_1$-ATPase that exhibits rotational motions to allow proteins or other materials to pass through the membrane.
These rotary enzymes are classified as membrane proteins because they are embedded in biological membranes to be responsible for various life-sustaining processes.

One of the characteristics of these motor proteins is that they consume chemical energy in order to deliver mechanical work such as unidirectional movements.
Each mechanical step is related to the free energy of ATP hydrolysis $\Delta E$, and one can roughly estimate the force $f$ exerted by a motor protein as~\cite{phillips2012}
\begin{align}
f = \frac{\Delta E}{\ell}\approx \frac{20\,k_{\rm B}T}{8\,{\rm nm}} \approx 10\,{\rm pN},
\label{eq:force}
\end{align}
where $\ell\approx8$\,nm is the typical distance traveled by a kinesin motor and $k_{\rm B}T\approx4\times10^{-21}$\,J with $k_{\rm B}$ and $T$ being the Boltzmann constant and the temperature, respectively.
Furthermore, by assuming that motor proteins or enzymes consist of an elastic spring, one can estimate the motor's elastic constant $k$ and the characteristic relaxation timescale $\tau$ as
\begin{align}
k=\frac{f}{a}\approx\frac{10\,{\rm pN}}{10\,{\rm nm}}\approx10^{-3}\,{\rm N/m},
\end{align}
and
\begin{align}
\tau = \frac{\zeta}{k}\approx\frac{10^{-7}\,{\rm N}\cdot{\rm s/m}}{10^{-3}\,{\rm N/m}}\approx10^{-4}\,{\rm s},
\end{align}
respectively.
Here, we have chosen the protein size as $a\approx10$\,nm and the friction coefficient of a motor as
$\zeta\approx10^{-7}$\,${\rm N}\cdot{\rm s/m}$.
These estimates give characteristic physical quantities for motor proteins at small scales.

Since their mechanical work becomes available by attaching themselves to some biological structures such as filaments or membranes, motile behavior was not reported for enzymatic molecules that are freely dispersed in aqueous solutions or cytoplasm and are not attached to surrounding structures.
However, it has been experimentally shown that enzymes also exhibit mechanical motions, and their dependency on substrate concentrations or theoretical modelings has attracted much attention, as we shall explain below.

\subsection{Biocatalysis by enzymatic molecules}

Enzymes are functional macromolecular proteins consisting of amino acids in a particular sequence~\cite{albertsbook}.  
The assembly of amino acids folds into a precise 3D conformation with reactive sites on its surface.
Therefore, these amino acid polymers bind with high specificity to other molecules and act as enzymes catalyzing chemical processes that make or break covalent bonds of other molecules~\cite{albertsbook,phillips2012}.
Moreover, these proteins play other roles, such as maintaining structures, generating movements, and sensing signals, which are essential for cellular metabolism and homeostasis~\cite{albertsbook,phillips2012}.

In order to exhibit specific functions in cells, the shapes of most biological macromolecules need to be highly constrained~\cite{albertsbook}.
However, most of the covalent bonds in a macromolecule allow rotation of atoms and give the polymer chain great flexibility.
This polymer chain flexibility allows a macromolecule to adopt an almost unlimited number of conformations caused by random thermal motions of surrounding environments~\cite{albertsbook}. 
Macromolecules can fold tightly into highly preferred conformations due to many weak noncovalent bonds between different parts of the same molecule~\cite{albertsbook}.

The four types of noncovalent interactions (hydrogen bonds, van der Waals attractions, hydrophobic forces, and electrostatic attractions) are essential for biological molecules~\cite{albertsbook}.
Although the strength of these noncovalent bonds is $20$ times weaker than that of a covalent bond, they provide tight binding once many of such weak interactions are formed simultaneously~\cite{albertsbook}.
In addition, they can also add up to create a strong attraction between two different molecules when they fit together very closely~\cite{albertsbook}.
Since the strength of the binding depends on the number of noncovalent bonds formed between molecules, interactions of almost any affinity are possible~\cite{albertsbook,phillips2012}.
This flexibility in noncovalent bonds causes rapid molecule dissociation, which drives catalytic chemical cycles followed by the dissociation of product molecules~\cite{albertsbook,phillips2012}.

Recent advances in fluorescence microscopy have motivated studies of single-molecule observation.
In 1988, for example, Lu \textit{et al.}\ investigated enzymatic turnovers of single cholesterol oxidase molecules in real-time by monitoring the emission from the enzyme fluorescent active site~\cite{lu1998}.
Moreover, they derived the waiting time distribution of the enzymes and showed that the obtained distribution agrees well with that derived from real-time trajectories of enzymes~\cite{lu1998}.
Later, it was shown that the reaction velocity for single-enzyme observations coincides with that for ensemble-enzyme observations 
when scaled by the total concentration of enzymes~\cite{kou2005,english2006}.
This coincidence results from the fact that the average over the long time trace of a single molecule is equivalent to that over a large ensemble of identical molecules.

\subsection{Conformational dynamics during chemical reactions}

\begin{figure}[tbh]
\centering
\includegraphics[scale=.9]{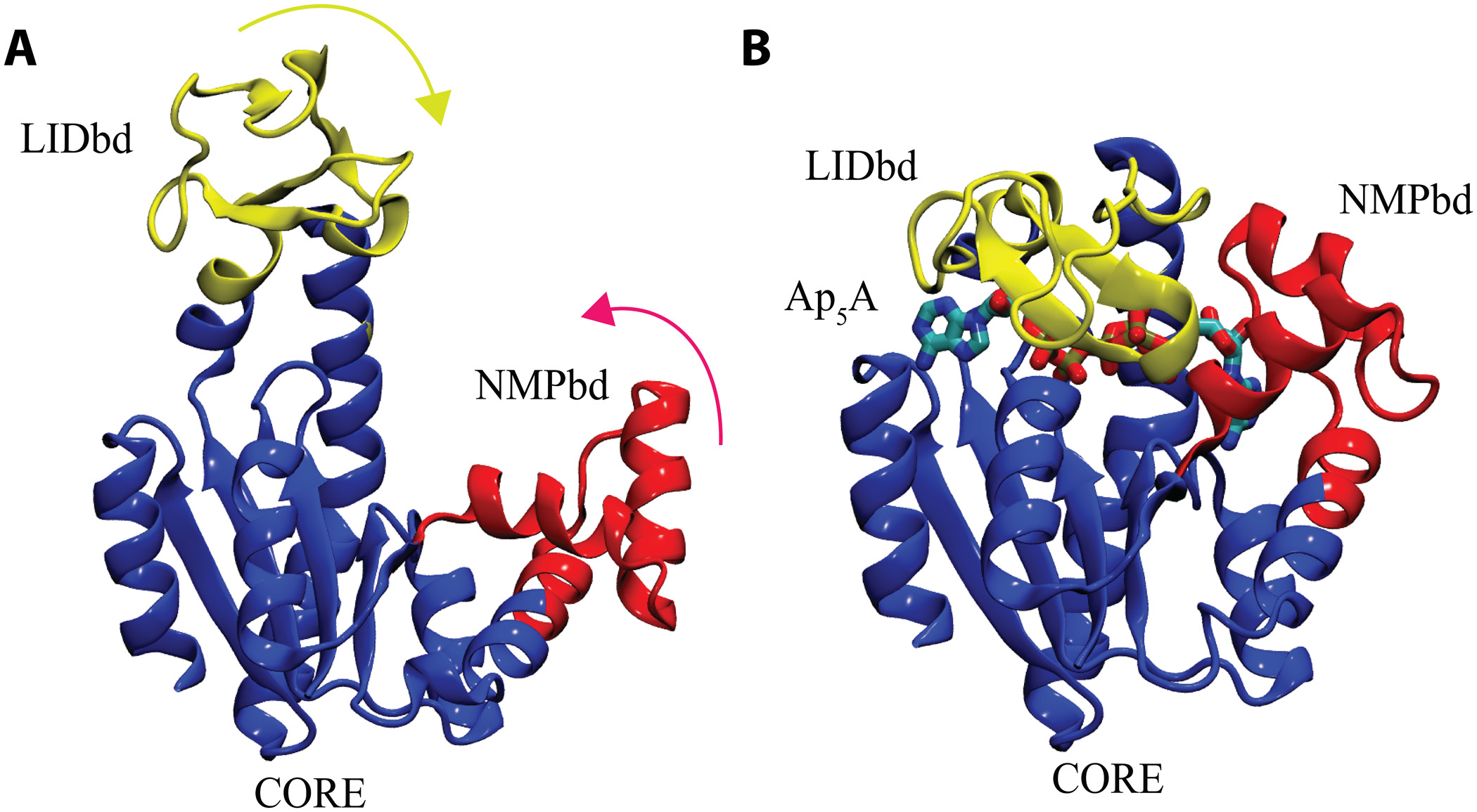}
\caption{
Conformational changes of adenylate kinase from the fully open (A) to fully closed (B) conformations in the sequential binding mechanism.
Adapted from [F.\ Or\"{a}dd, H.\ Ravishankar, J.\ Goodman, P.\ Rogne, L.\ Backman, A.\ Duelli, M.\ N.\ Pedersen, M.\ Levantino, M.\ Wulff, M.\ Wolf-Watz, and M.\ Anderson, Sci.\
Adv.\ \textbf{7}, eabi5514 (2021)]~\cite{oradd2021tracking} under CC BY-NC 4.0.
}\label{fig:ADKchange}
\end{figure}

Motor proteins such as myosin and kinesin undergo unidirectional motion responsible for autonomously contracting muscles and transporting materials within cells~\cite{albertsbook}.
By catalyzing ATP hydrolysis, they acquire sufficient energy to exhibit the motile behavior.
On the other hand, most  enzymes do not exhibit such behavior, although both motor proteins and enzymes catalyze chemical reactions.
In particular, enzymes exhibit a distinctive type of dynamics, i.e., conformational change, generally induced by substrate binding and product release~\cite{gerstein1994,oradd2021tracking}.
Then, it follows that enzymes undergo a conformational change in each turnover cycle of the chemical reactions in the presence of substrate molecules, as shown in 
Fig.~\ref{fig:ADKchange}.

These conformational dynamics have been taken into account to mimic  molecular enzymes in a framework of the elastic network model.
Togashi \textit{et al.}\ analyzed nonlinear conformational relaxation dynamics of proteins in elastic networks and found that motions of these proteins are robust against external perturbations~\cite{togashi2007}.
They also constructed an example of an artificial elastic network, operating as a cyclic machine powered by substrate binding with the use of evolutionary optimization methods.
Later, Echeverria \textit{et al.}\ presented a multi-scale coarse-grained description of protein conformational dynamics in a solvent, which is described by multiparticle collision dynamics~\cite{echeverria2011}.
They found that hydrodynamic interactions cause essential effects on the large-scale conformational motions of the protein and significantly affect the translational diffusion coefficients and orientational correlation times.

Recently, using direct observation techniques, Aviram \textit{et al.}\ studied the relationship between conformational dynamics and the chemical steps of enzymes~\cite{aviram2018}.
They labeled adenylate kinase, which is responsible for cellular energy homeostasis, from \textit{E.\ coli} with FRET dyes at CORE and LID domains.
Then they derived open and closed conformations of the enzyme from histograms of FRET efficiency.
The obtained histograms show a peak FRET efficiency value of $0.4$ in the absence of substrates, while the peak shifts to $0.6$ in the saturating concentrations of ATP.
By comparing these conformational dynamics and chemical steps, they found that substrate binding increases domain closing and opening times dramatically, i.e., $100$--$200$ times faster than the enzymatic turnover rate~\cite{aviram2018}.

ATP synthase or F$_0$F$_1$-ATPase is composed of two different rotary motors (F$_0$ and F$_1$) connected to a shaft and shows another type of conformational change, i.e., rotational motion.
The F$_0$ motor uses the gradient of hydrogen ions to rotate, while the F$_1$ motor uses APT hydrolysis to rotate in the opposite direction of F$_0$~\cite{phillips2012}.
When the transmembrane electrochemical gradient is strong, F$_0$ generates more torque than F$_1$, and it rotates in reverse to synthesize ATP~\cite{phillips2012}.
When the electrochemical gradient is weak, on the other hand, the torque that F$_1$ generates dominates over that of F$_0$, and the APT hydrolysis occurs, which pumps hydrogen ions out of the cell~\cite{phillips2012}.
By direct observation of the motion of F$_1$, Noji \textit{et al.}\ showed that the motor rotates in distinct steps of $120^\circ$, and the induced torque is the order of $10$\,pN$\cdot$nm~\cite{noji1997,albertsbook}.
Given the nanometer-size rotary motor, one can see that the observed torque is comparable to the force exerted by a motor protein, as we have estimated in Eq.~(\ref{eq:force}).

\subsection{Diffusion enhancement in enzyme solutions}

To explicitly focus on enzyme-driven phenomena, diffusion in enzyme solutions has been 
experimentally studied in recent years~\cite{muddana2010,sengupta2013,riedel2015,illien2017_2,zhao2017,jee2018,dey2016,wang2020}.
For example, Muddana \textit{et al.}\ first reported the enhanced diffusion of enzyme urease in the presence of 
substrate urea~\cite{muddana2010}.
Later, it was shown that enzymes exhibit collective motions toward the direction of higher or lower 
concentrations of substrates, i.e., chemotaxis and antichemotaxis, respectively~\cite{sengupta2013,jee2018,wang2020}.

It was also claimed that the enhanced diffusion had been observed even during catalysis at the {\AA}ngst\"{o}m scale, which is much smaller than a system of molecular enzymes~\cite{dey2016,wang2020}.
Since the used catalyst shows less conformational dynamics due to its larger rigidity compared to molecular proteins, a different mechanism, such as the transfer of momentum from the active catalyst molecule, was proposed to account for the enhanced diffusion.
Moreover, it was also reported that enhanced diffusion in molecular-scale systems was due to a convective flow~\cite{macdonald2019}, and enhancement in diffusivity is still a matter of debate~\cite{rezaei2022}.

To identify the mechanism of the observed enhanced diffusion and chemotactic phenomena, several people have suggested theories that account for the roles of heat or hydrodynamic interaction caused by 
enzymes~\cite{mikhailov2015,kapral2016,golestanian2015,illien2017}.
Mikhailov \textit{et al.}\ discussed the collective hydrodynamic flows induced by active force dipoles and analytically derived the diffusion enhancement of a tracer particle, which depends linearly on the enzyme activity~\cite{mikhailov2015,kapral2016}.
Then, Golestanian proposed four mechanisms for the enhanced diffusion of enzymes, namely, self-thermophoresis, boost in kinetic energy, stochastic swimming, and collective heating~\cite{golestanian2015}.
He concluded that only the last two descriptions could account for the phenomenon.
Later, Illien \textit{et al.}\ took into account the hydrodynamic effects induced by conformational changes of enzymatic domains and demonstrated that a single enzyme could diffuse faster even at equilibrium~\cite{illien2017}.
However, a recent experiment pointed out the difficulty of quantitatively accounting for the observed 
enhanced diffusion within the suggested theoretical approaches~\cite{xu2019}.

\section{Nonequilibrium Transport in Living Systems}
\label{sec:nonequilibrium}

\subsection{Active transport in biological cells}

\begin{figure*}[tbh]
\begin{center}
\includegraphics[scale=0.28]{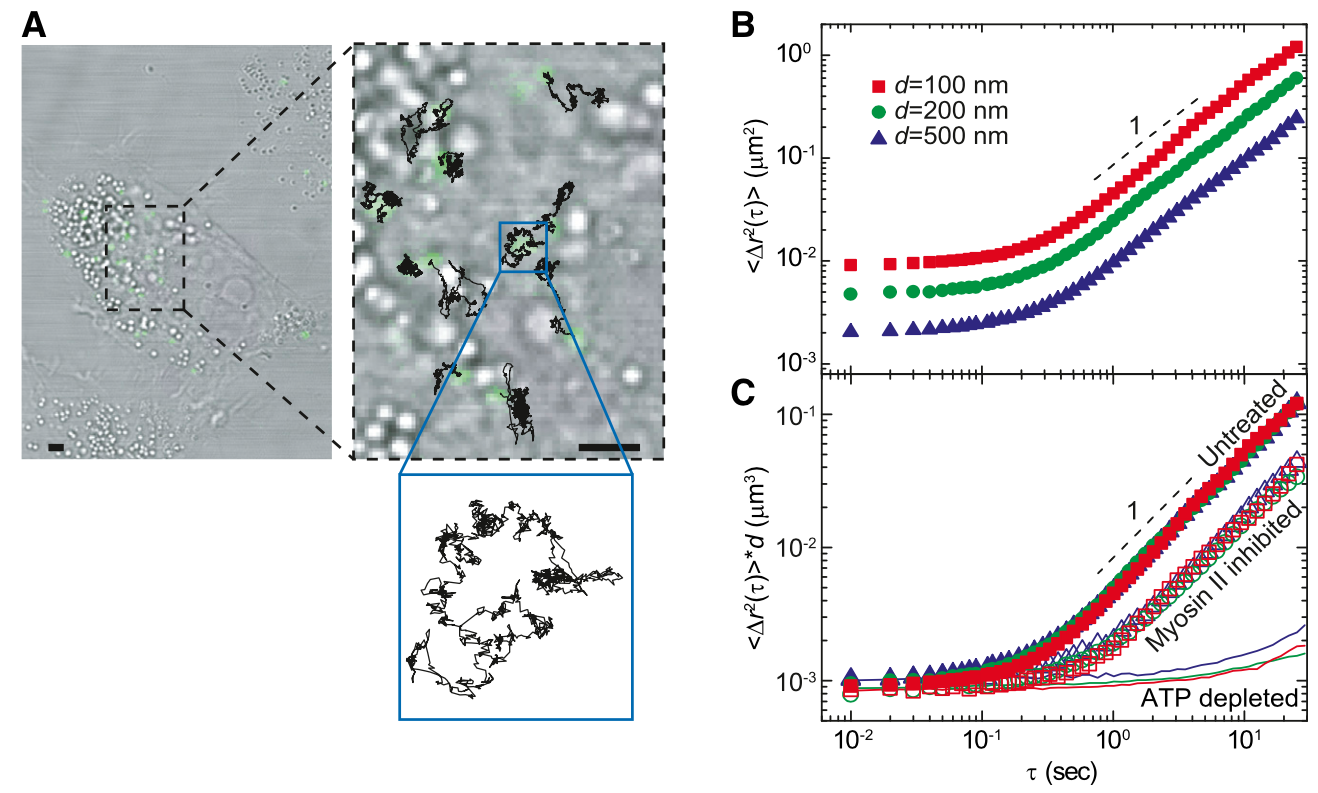}
\end{center}
\caption{
(A) Bright-field image of an A7 cell with microinjected 200\,nm-diameter fluorescence particles (green) and 2\,min trajectories (black) superimposed on top.
Scale bar, 5\,$\mu$m.
(B) Two-dimensional ensemble-averaged mean-square displacement of tracer particles of various sizes are plotted against lag time on a log-log scale in living A7 cells.
Red, green, and blue symbols and lines represent particles that are 100, 200, and 500\,nm in diameter, respectively.
(C) Ensemble-averaged mean-square displacement scaled with particle diameter in untreated (solid symbols), blebbistatin treated (open symbols), and ATP-depleted (solid lines) A7 cells.
Reprinted from Cell, \textbf{158}, 822, M.\ Guo, A.\ J.\ Ehrlicher, M.\ H.\ Jensen, M.\ Renz, J.\ R.\ Moore, R.\ D.\ Goldman, J.\ Lippincott-Schwartz, F.\ C.\ MacKintosh, and D.\ A.\ Weitz, ``Probing the Stochastic, Motor-Driven Properties of the Cytoplasm Using Force Spectrum Microscopy'', p.\ 823, Copyright (2014), with permission from Elsevier~\cite{guo2014}.
\label{fig:Guo}
}
\end{figure*}

In recent years, to better understand nonequilibrium phenomena in biological cells, the diffusive properties 
of tracer particles \textit{in vivo} have been experimentally studied ~\cite{guo2014,parry2014}.
For example,
Guo \textit{et al.}\ microinjected submicron colloidal particles into A7 melanoma cells and measured their time-dependent motion to calculate ensemble-averaged mean-square displacement, as shown in 
Fig.~\ref{fig:Guo}(A)~\cite{guo2014}.
In Fig.~\ref{fig:Guo}(B), the mean-square displacement shows constant behavior at small timescales, whereas, at large timescales, it increases approximately linearly with time~\cite{guo2014}.
Although this linearly increasing behavior is consistent with Brownian motion in a purely viscous liquid at thermal equilibrium, such a description can not be applied to the cytoplasm.
They also observed the mean-square displacement in cells whose activity is inhibited and found no change in the displacement compared to that in active cells at small timescales, as shown in Fig.~\ref{fig:Guo}(C)~\cite{guo2014}.
At large timescales, on the other hand, the mean-square displacement exhibits an increasing behavior when myosin is inhibited or a nearly time-independent behavior when ATP is depleted~\cite{guo2014}.

These results suggest that motor proteins and ATP-driven proteins such as enzymes play essential roles in the motion of particles in cells.
Parry \textit{et al.}\ performed similar experimental studies in the bacterial cytoplasm and observed the enhanced diffusion of plasmids in untreated cells~\cite{parry2014}.
Such a phenomenon is called anomalous diffusion to distinguish from the ordinary Brownian diffusion that is induced by the thermal motions of solvent molecules.

The above experimental findings demonstrate that passive particles diffuse faster when cells operate properly in the presence of substrate molecules.
They also imply that nonequilibrium fluctuations due to continuous energy supply contribute to nonthermal diffusion.
At the same time, ATP-dependent diffusion observed in the cytoplasm suggests that enzymes that catalyze chemical reactions using substrate molecules also have contributions to anomalous diffusion.
However, due to the complexity of cellular environments that contain structures and materials such as cytoskeletons and viscoelastic media, the mechanism of anomalous diffusion has not yet been completely clarified.

\subsection{Rheology of cellular systems}

Biomolecular machines exhibit mechanical motions in fluid environments such as cytoplasm or biological membranes, and the physical properties of the fluid with these active constituents are essential for biomolecular transports and chemical reactions~\cite{saintillan2018}.
Hence, the effect of enzymatic activity on the rheological properties of such active fluids has gathered much attention in recent years.

Before reviewing some experimental findings in this field, we first mention the concept of the rheological properties of ordinary passive fluids.
In general, the rheological properties of fluids are characterized by the fourth-rank viscosity tensor $\eta_{ijk\ell}$ that linearly connects the strain rate tensor $v_{ij}=(\partial_iv_j+\partial_jv_i)/2$ and the fluid stress tensor $\sigma_{ij}$~\cite{Landau1987}:
\begin{align}
\sigma_{ij} = \eta_{ijk\ell}v_{k\ell},
\end{align}
where the indices $i,j,k,\ell=x,y,z$, and we assume summation over repeated indices.
In the above, $\mathbf{v}$ is the fluid velocity field, and the viscosity tensor for a 3D isotropic fluid is given by~\cite{Landau1987}
\begin{align}
\eta_{i j k \ell}=& \eta_{\mathrm{d}} \delta_{i j} \delta_{k \ell}
+\eta\left(\delta_{i k} \delta_{j \ell}+\delta_{i \ell} \delta_{j k}-\frac{2}{3}\delta_{i j} \delta_{k \ell}\right),
\label{eq:viscosity}
\end{align}
where $\delta_{ij}$ is the Kronecker delta, and $\eta_{\mathrm{d}}$ and $\eta$ are the dilatational and shear viscosities, respectively.
The shear viscosity can be obtained by the autocorrelation functions of the viscous stress $\sigma_{xy}$ based on the linear response theory 
\begin{align}
\eta = \frac{1}{k_{\rm B}TV}\int_0^\infty dt\, \langle \sigma_{xy}(t) \sigma_{xy}(0) \rangle,
\end{align}
where $V$ is the volume and $\langle\cdots\rangle$ denotes the average over the steady-state ensemble of trajectories.
For a passive fluid, its viscosities can be modified, e.g., by the density of the immersed particle, the system temperature, or applied shear forces~\cite{saintillan2018}.
On the other hand, these rheological properties can be modified by additional contributions to the cytoplasm or biological membranes where metabolic activities are present, and the systems are strongly driven out of equilibrium.

\begin{figure}[tbh]
\begin{center}
\includegraphics[scale=0.15]{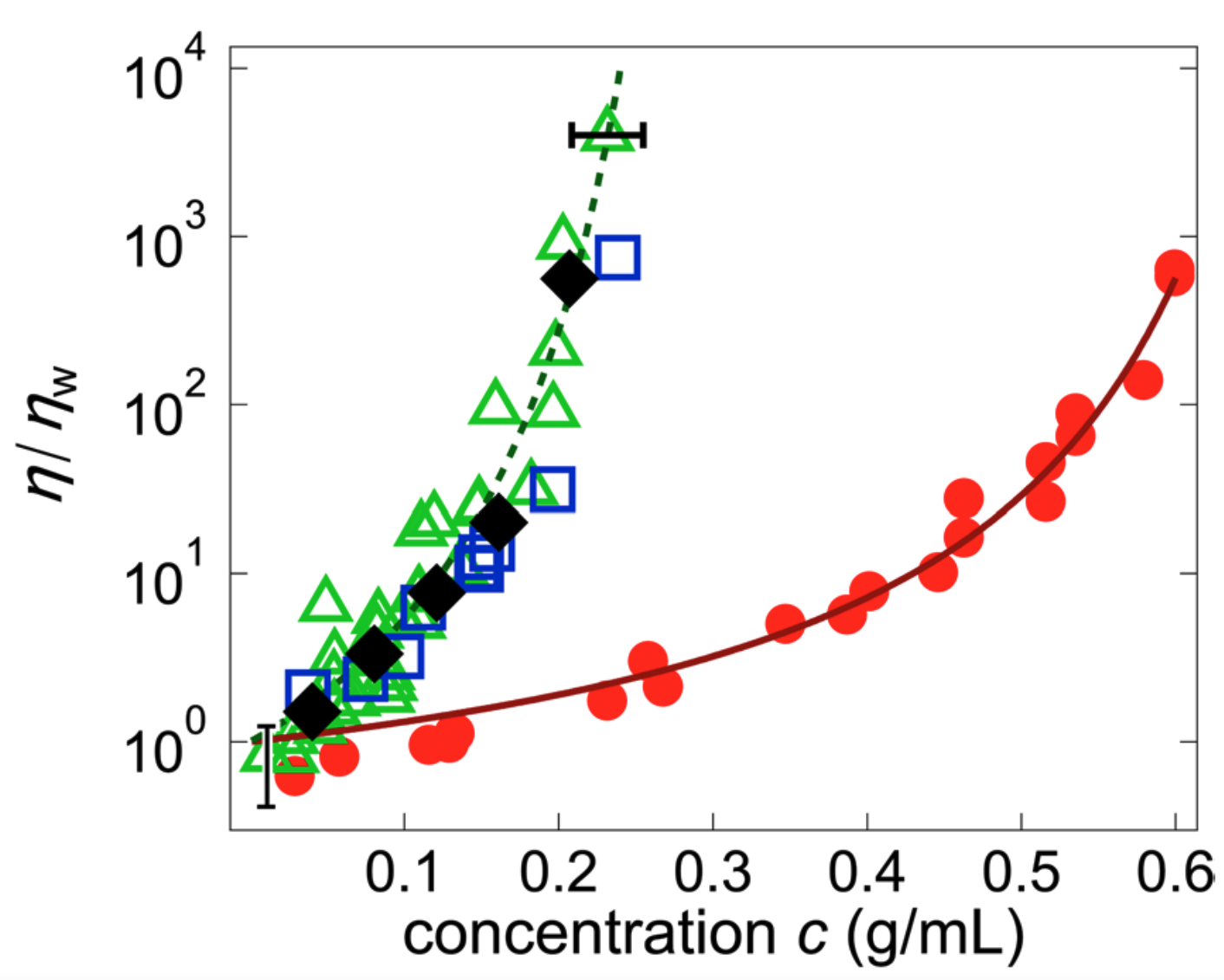}
\end{center}
\caption{
The effective viscosity $\eta$ of BSA solutions (red circles) and cell extracts without cytoskeletons and metabolic activity (green triangles: \textit{E. coli}, blue squares: Xenopus eggs, and black diamonds: HeLa cells) as a function of the concentration of the macromolecules $c$.
The viscosity $\eta$ is rescaled by the water viscosity $\eta_{\rm w}$.
Adapted from [K.\ Nishizawa, K.\ Fujiwara, M.\ Ikenaga, N.\ Nakajo, M.\ Yanagisawa, and
D.\ Mizuno, Sci.\ Rep.\ \textbf{7}, 1 (2017)]~\cite{nishizawa2017} under CC BY 4.0.
\label{fig:nishizawa}
}
\end{figure}

\begin{figure}[tbh]
\centering
\includegraphics[scale=.08]{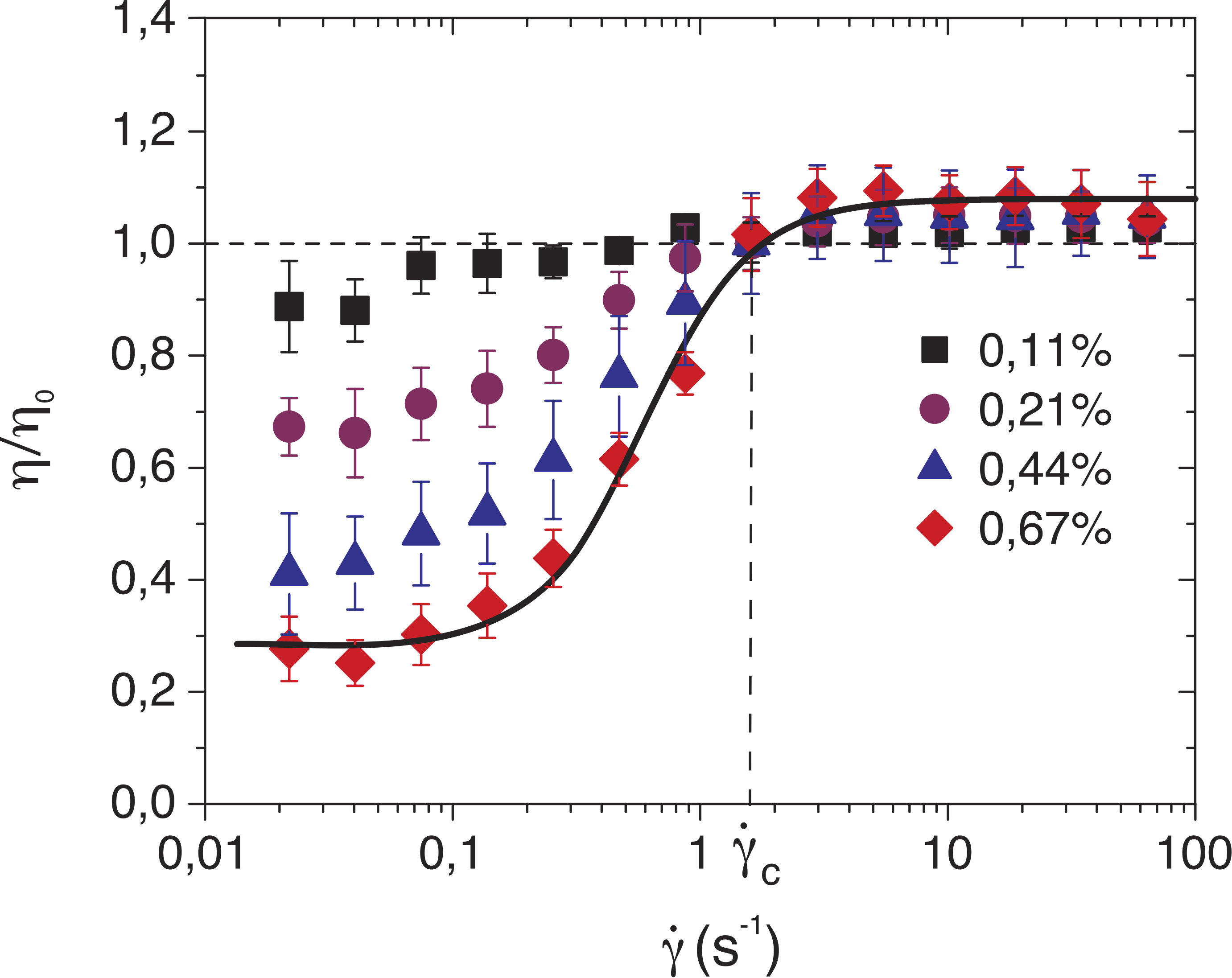}
\caption{
Effective viscosity of \textit{E.\ coli} suspensions as a function of the applied shear rate $\dot{\gamma}$ for various values of the volume fraction $\phi$.
Reprinted figure with permission from [H.\ M.\ L\'{o}pez, J.\ Gachelin, C.\ Douarche, H.\ Auradou, and E.\ Cl\'{e}ment, Phys.\ Rev.\ Letts.\ \textbf{115}, 028301 (2015)]~\cite{lopez2015}.
Copyright (2015) by the American Physical Society.
\label{fig:lopez2015}
}
\end{figure}

Nishizawa \textit{et al.}\ experimentally studied the shear viscosity of cytoplasm for various concentrations of macromolecules~\cite{nishizawa2017}.
The viscosity of cell extracts without metabolic activation rapidly increased with the macromolecule concentration, which shows diverging viscosity at the critical concentration $c^\ast\approx 0.34$\,g$/$mL, as shown in Fig.~\ref{fig:nishizawa}.
This critical concentration is close to the physiological concentration in living cells ($\sim0.3$\,g$/$mL).
On the other hand, metabolically active living cells showed moderate fluidity and did not undergo glass transition unlike inactivated cells~\cite{nishizawa2017}.
These experimental findings suggest that metabolically active living cells are essential for finite fluidity that facilitates the efficient transport of molecules in living cells.

In more macroscopic scales, the shear viscosity of bacterial suspensions has been studied~\cite{saintillan2018,rafai2010,lopez2015}.
First, Rafa\"{i} \textit{et al.}\ performed experiments on the rheology of suspensions of live cells, \textit{Chlamydomonas Reinhardtii}, and revealed that the obtained viscosity was larger than that of suspensions with the same volume fraction of dead cells~\cite{rafai2010}.
Later, L\'{o}pez \textit{et al.}\ investigated the response of an \textit{E.\ coli} suspension under the shear flow and showed that the suspension viscosity decreases with increasing the bacterial density at low shear rates as 
shown in Fig.~\ref{fig:lopez2015}~\cite{lopez2015}.
These experimental findings suggest that active constituents that convert chemical energy into mechanical work contribute to rheological properties and make them dependent on the internal activity~\cite{saintillan2018}.

\subsection{Emergent patterns in active chiral systems}

In addition to the above peculiar rheological properties, the emergent macroscopic patterns have been investigated in active fluids~\cite{sumino2012,tabe2003,yamauchi2020,beppu2021}.
Experimentally, such active systems have been realized in nanoscale molecular motors~\cite{sumino2012,tabe2003} or multicellular biological systems~\cite{yamauchi2020,beppu2021}.
Sumino \textit{et al.}\ investigated the behavior of microtubules propelled by surface-bound dyneins and observed that self-organization of the microtubules results in vortices at high densities due to the alignment mechanism~\cite{sumino2012}.
In addition, a spatiotemporal pattern was found in the monolayer of synthetic molecular motors~\cite{tabe2003}.
For larger scales such as multicellular systems, edge flows were observed at the boundary of active nematic cells~\cite{yamauchi2020}.
In the bacterial suspensions, Beppu \textit{et al.}\ showed that edge currents develop as the bacterial density increases~\cite{beppu2021}.

The common feature of these emergent chiral patterns is that the parity symmetry is broken due to the collective effects of motor proteins~\cite{sumino2012}, the chiral structure of molecules~\cite{tabe2003}, or the surrounding geometries~\cite{yamauchi2020,beppu2021}.
Moreover, these active constituents continuously consume energy, and hence the time-reversal symmetry is broken.
Since both the time-reversal and parity symmetries are violated in these biological environments, they are called active chiral systems or 
active chiral fluids~\cite{banerjee2017}.
In these out-of-equilibrium systems, the equilibrium concept such as free energy, detailed balance, and time-reversal symmetry no longer holds~\cite{gompper2020}, and new physical quantities that characterize the systems are necessary, as we describe in the next Sections.

\section{Coarse-Grained Modeling of Biological Nanomachines}
\label{sec:coarse-graind}

\subsection{Continuum hydrodynamic description}
\label{subsec:continuum}

For the passive case without any activity, the disturbance flow arises only when an external force or flow field is imposed on a fluid~\cite{saintillan2018}.
For the active case, however, the disturbance flow is induced even in a quiescent fluid because of the mechanical work driven by motor proteins or enzymatic molecules.
Over the length scale of molecular proteins where the inertial effect is negligible, the hydrodynamic behavior is governed by the well-known Stokes equation
\begin{align}
\eta \nabla^2\mathbf{v}(\mathbf{r}) -\nabla p(\mathbf{r})   +\mathbf{F}(\mathbf{r}) =0, 
\label{eq:stokes}
\end{align}
and the incompressibility condition
\begin{align}
\nabla\cdot\mathbf{v}(\mathbf{r})=0,
\label{eq:incompressible}
\end{align}
where $\nabla=(\partial_x,\partial_y,\partial_z)$ is 
the 3D differential operator and $\mathbf{r}=(x,y,z)$  
is the position vector.
In the above, $\eta$ is the shear viscosity, $\mathbf{v}$ is the fluid velocity field, $p$ is the hydrostatic pressure, and $\mathbf{F}$ is any other arbitrary force density on the fluid.

\begin{figure*}[tbh]
\centering
\includegraphics[scale=0.3]{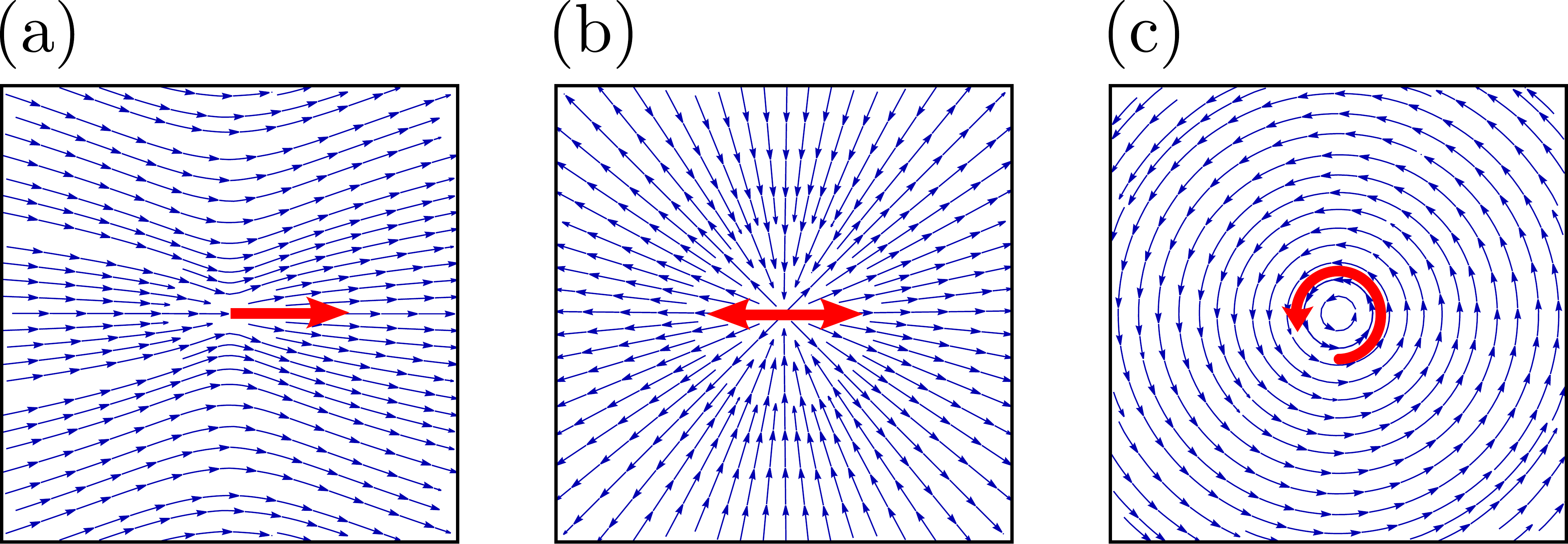}
\caption{
Streamlines induced by (a) Stokeslet for $n=0$ and (b) stresslet and (c) rotlet for $n=1$ in Stokes flows.
\label{fig:oseen}
}
\end{figure*}

In this Stokes regime, the hydrodynamic flow at $\mathbf{r}$ induced by $\mathbf{F}$ at $\mathbf{r}_0$ is expressed as $v_i(\mathbf{r})=-\int d\mathbf{r}_0\, G_{ij}\left(\mathbf{r}-\mathbf{r}_{0}\right) F_j\left(\mathbf{r}_{0}\right)$, where the Green's function 
(propagator) for a 3D unbounded fluid, namely, the Oseen tensor is~\cite{doi1988}
\begin{align}
G_{ij}(\mathbf{r}) = \frac{1}{8\pi\eta r} \left( \delta_{ij} + \frac{r_ir_j}{r^2} \right),
\label{eq:oseen}
\end{align}
with $r=|\mathbf{r}|$.
If the position of $\mathbf{r}$ is far from that of $\mathbf{F}$, a Taylor expansion of the Green's function provides a far-field representation of the flow in terms of multipole moments $\mathbf{M}^{(n)}$ of the tractions and velocities~\cite{saintillan2018}.
Then the velocity field can be expressed in terms of $\mathbf{G}$ and its derivatives as~\cite{saintillan2018,kim2013}
\begin{align}
v_i(\mathbf{r}) \approx \sum_{n=0}^{\infty} \partial_k^{(n)} G_{ij}(\mathbf{r})M_{jk}^{(n)}.
\label{eq:multipole}
\end{align}
Figure~\ref{fig:oseen} represents the streamlines induced 
by the Stokeslet $\mathbf{M}^{(0)}=\mathbf{F}$ when $n=0$, and also the symmetric (stresslet, $\mathbf{S}$) and antisymmetric (rotlet, $\mathbf{L}$) parts of $\mathbf{M}^{(1)}$ when $n=1$.

In a biological context, no force and torque act on nanomachines as they function autonomously in the presence of chemical energy, requiring the condition $\mathbf{F}=\mathbf{L}=0$ to hold in general.
Hence, the stresslet $\mathbf{S}$ and the torque dipole $\mathbf{M}^{(2)}$ have been used to model enzymatic molecules~\cite{mikhailov2015,kapral2016} and rotary proteins~\cite{oppenheimer2019,lenz2003}, respectively.
For biological membranes, moreover, one has to derive the mobility tensor for free~\cite{saffman1975,saffman1976}, confined~\cite{evans1988}, or curved~\cite{henle2010} geometries of a 2D fluid, which can be derived from the 2D version of Eq.~(\ref{eq:stokes})~\cite{oppenheimer2009,oppenheimer2010,ramachandran2011}.

\subsection{Active force dipole model}

\begin{figure*}[tbh]
\centering
\includegraphics[scale=0.15]{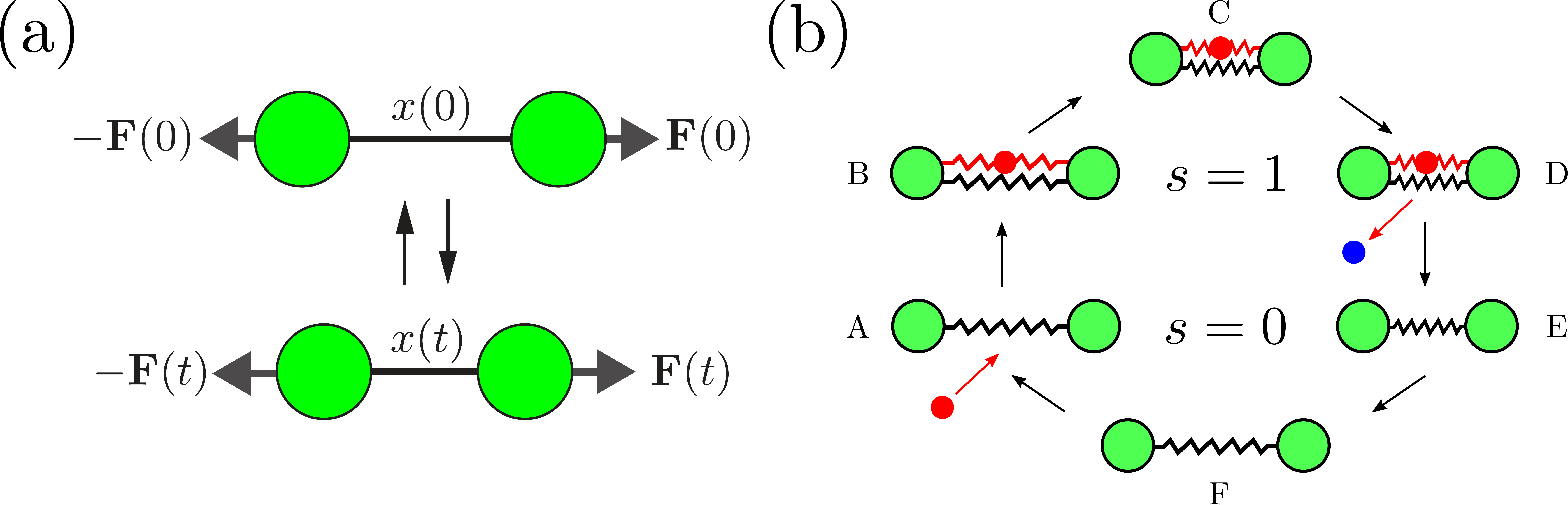}
\caption{
(a) The active force dipole for a molecular enzyme that undergoes the conformational change cyclically in the presence of substrate molecules.
In the model, the time-dependent distance and the force of the enzyme are $x(t)$ and $\mathbf{F}(t)$, respectively.
(b) The turnover cycle and mechanochemical motions in the active dimer model of an enzyme.
Adapted from [Y.\ Hosaka, S.\ Komura, and A.\ S.\ Mikhailov, Soft Matter \textbf{16}, 10734 (2020)]~\cite{hosaka2020_2} under CC BY 3.0.
\label{fig:dipole_intro}
}
\end{figure*}

Next, we shall present more detailed descriptions of the active force dipole model that was originally proposed by Mikhailov and Kapral~\cite{mikhailov2015,kapral2016}.
As explained in Sec.~\ref{sec:intro}, each actual enzyme has a specific 3D conformation that depends on its biological function and the surrounding environments, such as the cytoplasm and biological membranes.
At large scales, however, any enzyme can be regarded as an active force dipole, as shown in Fig.~\ref{fig:dipole_intro}(a).
The active force dipole consists of two domains, representing enzymatic domains, connected with a shaft, and its length cyclically varies in time to mimic the conformational dynamics of enzymes during chemical reactions.

Since a dipole exerts the time-dependent force, $\mathbf{F}(t)$, along its axis direction, the dipole induces the hydrodynamic flow in surrounding environments.
For example, if a force dipole is immersed in a 3D fluid, the generated flow field can be calculated from Eq.~(\ref{eq:multipole}) for $n=1$: 
\begin{align}
v_{i}(\mathbf{r})=-F(t)x(t) \hat{x}_{k} \partial_{k} G_{i j}(\mathbf{r}) \hat{x}_{j},
\end{align}
where $\mathbf{F}(t)=F(t)\hat{\mathbf{x}}$ and $\hat{\mathbf{x}}$ being a unit vector in the direction of the enzyme axis.
In the context of self-propelled microswimmers, $F>0$ $(F<0)$ denotes a pusher (puller) type of a micromachine.
The resultant flow is the stresslet that is shown in Fig.~\ref{fig:oseen}(b).

When multiple dipoles are immersed in fluids, they induce collective hydrodynamic flows in their surroundings, which can lead to nonthermal fluctuations.
Considering these hydrodynamic effects, Mikhailov \textit{et al.}\ derived the diffusion coefficient of a passive tracer in a solution where dipoles are homogeneously distributed in space, and the directions of their long axes are randomly distributed.
Moreover, when dipole concentration gradient is present, the tracer exhibits chemotaxis, which was observed in the experiment~\cite{sengupta2013}.
Later, Koyano \textit{et al.}\ discussed the situation where dipoles are aligned and concentrated in a liquid domain corresponding to lipid rafts in biological membranes~\cite{koyano2016}.
Hosaka \textit{et al.}\ considered the hydrodynamic coupling between the 2D and 3D fluids and derived the active diffusion coefficients for an arbitrary size of the diffusing particle~\cite{hosaka2017}.

The hydrodynamic interactions and clustering mechanisms of active force dipoles were also investigated in flat~\cite{manikantan2020} or curved~\cite{bagaria2021} biological membranes.
Manikantan examined the phase behavior of a pair of hydrodynamically interacting force dipoles and showed that bulk confinement plays an essential role in the clustering dipoles.
Moreover, it was demonstrated that multiple dipoles exhibit the collective dynamics that can be tuned by the confinement of the membrane.
In curved geometry, aggregation effects of dipoles were confirmed in the regimes of both low and high curvatures~\cite{bagaria2021}.
One of the unique features of 2D fluid membrane geometries is the existence of hydrodynamic screening lengths that make the short-distance hydrodynamic behavior significantly different from the long-distance one~\cite{saffman1975,saffman1976,evans1988}.

Mechanochemically active enzymes change their shapes within every turnover cycle, as shown in Fig.~\ref{fig:dipole_intro}(b).
Therefore, they induce circulating flows in the solvent around them and behave as oscillating hydrodynamic force dipoles.
Because of nonequilibrium fluctuating flows that are collectively generated by the enzymes, mixing in the solution and diffusion of passive particles are expected to get enhanced. 
Hosaka \textit{et al.}\ investigated the intensity and statistical properties of such force dipoles in the minimal active dimer model of a mechanochemical enzyme~\cite{hosaka2020_2}.
In the framework of this model, estimates for collective hydrodynamic effects in solution and in lipid bilayers under rapid rotational diffusion were derived, and they examined
available experimental and computational data.

Later, Hosaka \textit{et al.}\ discussed the shear viscosity of a Newtonian solution of catalytic enzymes and substrate molecules~\cite{hosaka2020}.
The enzyme was modeled as a two-state dimer consisting of two spherical domains connected with an elastic spring. The enzymatic conformational dynamics are induced by the substrate binding, and such a process was represented by an additional elastic spring. 
Employing the Boltzmann distribution weighted by the waiting times of enzymatic species in each catalytic cycle, they obtained the shear viscosity of dilute enzyme solutions as a function of substrate concentration. The substrate affinity distinguishes between fast and slow enzymes, and the corresponding viscosity expressions were obtained. 
Furthermore, they connected the obtained viscosity with the diffusion coefficient of a tracer particle in enzyme solutions~\cite{hosaka2020}.

\subsection{Hydrodynamic coupling between domains}

So far, the hydrodynamic flow induced by a single or multiple enzymes has been discussed in terms of the active force dipole model~\cite{mikhailov2015,kapral2016}.
Next, we review some of the theories that account for the internal hydrodynamics between the domains of a single enzyme~\cite{illien2017,adeleke2019,golestanian2019,adeleke2019_2}.
Since an actual macromolecular enzyme is asymmetric in general, its internal degrees of freedom are coupled to its center of mass diffusion, which would lead to the change in the diffusion coefficient~\cite{golestanian2019}.

Considering the effect of conformational fluctuations of an asymmetric dumbbell model, Illien \textit{et al.}\ showed that thermal fluctuations could result in negative contributions to the overall diffusion coefficient~\cite{illien2017}.
In addition, the time dependence of the diffusion coefficient of a dumbbell was derived with the use of the path integral  formulation.
Later, Adeleke-Larodo \textit{et al.}\ studied the anisotropy effect  on the enzyme diffusive behavior and derived the long-time diffusion coefficient of an asymmetric dumbbell by using the moment expansion technique~\cite{adeleke2019}.
They also studied the response of an asymmetric dumbbell enzyme to an inhomogeneous substrate concentration and showed that the enzyme tends to align parallel or antiparallel to the gradient, depending on the enzyme affinity to the substrate~\cite{adeleke2019_2}.
Moreover, it was implied that the hydrodynamic interaction plays a vital role in the collective behavior of many interacting enzyme molecules~\cite{adeleke2019_2}.
These theoretical findings suggest that hydrodynamic interactions act between the enzyme and substrate molecules and lead to the diffusion enhancement of a single enzyme even at equilibrium states~\cite{illien2017,adeleke2019,adeleke2019_2,golestanian2019}.

\subsection{Nonreciprocity in active systems}

Active systems are driven strongly out of equilibrium because of the energy input that is continuously consumed by their constituents.
This implies the absence of equilibrium concepts such as free energy, detailed balance, time-reversal symmetry, and Newton's third law~\cite{gompper2020}.
The violation of Newton's third law means that interactions between the objects are nonreciprocal, a crucial feature of chemical interactions between two different species, e.g., synthetic catalytic colloids, biological enzymes, or whole cells and microorganisms~\cite{agudo2019}.
For a 3D fluid, the hydrodynamic interaction between two objects separated by distance $r$ is described by the Oseen tensor $G_{ij}(\mathbf{r})$ in Eq.~(\ref{eq:oseen}) as long as $r$ is large enough~\cite{doi1988}.
Under the exchange of the index $i\leftrightarrow j$, $G_{ij}(\mathbf{r})$ remains the same, and the symmetry relation, $G_{ij}=G_{ji}$, holds.
This is known as the reciprocal theorem in fluid dynamics~\cite{masoud2019}.
However, in active fluids driven by biological nanomachines, the reciprocal relations are expected to be violated, i.e., $G_{ij}\neq G_{ji}$, which leads to peculiar collective behavior in out-of-equilibrium systems.

\begin{figure*}[tbh]
\centering
\includegraphics[scale=0.3]{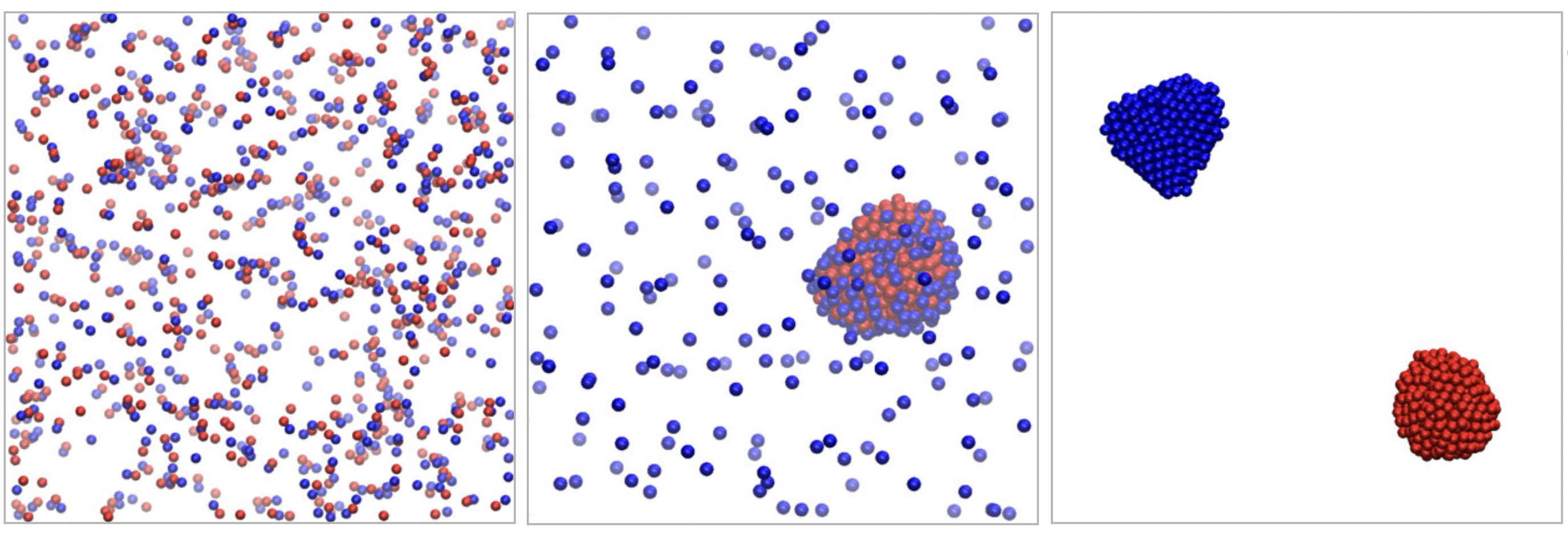}
\caption{
Binary mixtures of producer ($\alpha_1>0$, blue) and consumer ($\alpha_2<0$, red) species show (left) homogeneous states with the association of particles into small aggregation, (middle) a static dense phase that coexists with a dilute phase, and (right) separation into two static collapsed clusters.
Reprinted figure with permission from [J.\ Agudo-Canalejo and R.\ Golestanian, Phys.\ Rev.\ Letts.\ \textbf{123}, 018101 (2019)]~\cite{agudo2019}.
Copyright (2019) by the American Physical Society.
\label{fig:agudo2019}
}
\end{figure*}

Agudo-Canalejo and Golestanian theoretically studied mixtures of chemically interacting particles and demonstrated the existence of a new class of active phase separation phenomena where action-reaction symmetry or reciprocal relation is broken~\cite{agudo2019}.
Suppose that the concentration field of chemicals around a chemically active particle of species $i$ is  
$c\sim\alpha_i/r$, where $\alpha_i$ is the activity and 
$r$ is the distance to the particle's center, the motion of a particle of species $j$ in response to gradients of the chemical is given by a velocity $\mathbf{V}_{ij}\sim-\mu_j\nabla c=\alpha_i\mu_j\mathbf{r}_{ij}/r_{ij}^3$.
Here, $\mu_j$ is the mobility of the species $j$ and $\mathbf{r}_{ij}=\mathbf{r}_i-\mathbf{r}_{j}$ with $r_{ij}=|\mathbf{r}_{ij}|$.
Since the nonreciprocal relation, $\mathbf{V}_{ij}\neq\mathbf{V}_{ji}$, holds in general, an action-reaction symmetry is broken, which can not be seen at equilibrium states.
Moreover, such a nonreciprocity leads to various active phase separation phenomena, as shown in Fig.~\ref{fig:agudo2019},
for example.

Later, Ouazan-Reboul \textit{et al.}\ extended the above model by considering the size dispersity of the catalytically active particles and the dependence of catalytic activity on the substrate concentration~\cite{ouazan2021}.
In addition, a continuum model of pattern formation due to nonreciprocal interaction was proposed, and a traveling density wave was confirmed, which is a clear signature of broken time-reversal symmetry in such an active system~\cite{saha2020}.
However, despite these theoretical findings, studies on macroscopic physical quantities that lead to the emergence of the nonreciprocal relation in active systems are sparse, and further investigations are needed to estimate the extent of nonreciprocity in a biological context.

\section{Odd Viscosity in Nonequilibrium Systems}
\label{sec:odd}

\subsection{Origins of odd viscosity}

Odd viscosity is a rheological property that exists only when the time-reversal and parity symmetries are broken.
Although the concept was known for gasses or plasmas in an external magnetic field~\cite{lifshitz1981}, Avron \textit{et al.}\ showed in 1995 that the odd viscosity is present in a quantum Hall fluid and connected this viscosity with Berry curvature~\cite{avron1995}.
Since this study, the odd viscosity has been discussed in fractional quantum Hall and chiral super fluidic systems~\cite{read2009}.
Since the odd viscosity can be a new measure that characterizes a type of quantization or universality in these systems, this transport coefficient has gained much more attention in condensed matter and in active matter contexts that deal with living systems.

For a passive isotropic fluid, $\eta_{ijk\ell}$ is symmetric under the exchange of $i\leftrightarrow j$, while $\eta_{ijk\ell}=\eta_{ij\ell k}$ holds from the definition of the symmetric tensor $v_{k\ell}$, as can be inferred from Eq.~(\ref{eq:viscosity}).
Extending the above symmetry argument, Avron \textit{et al.} introduced a new type of index exchange, $ij\leftrightarrow k\ell$, which implies time-reversal transformation~\cite{avron1995,avron1998}.
For the passive case, the symmetry relation holds, i.e., $\eta_{ijk\ell}=\eta_{k\ell ij}$, as can be seen in Eq.~(\ref{eq:viscosity}), whereas the asymmetric (odd) part that satisfies $\eta_{{\rm o},ijk\ell}=-\eta_{{\rm o},k\ell ij}$ is a new contribution to the viscosity tensor.
For a 2D isotropic fluid, the odd part of the viscosity tensor can be written solely in terms of the scalar transport coefficient called \textit{odd viscosity} $\eta_{\rm o}$ as~\cite{ganeshan2017,epstein2020}
\begin{align}
\eta_{{\rm o},ijk\ell}= 
\frac{1}{2} \eta_{\rm o}\left(\epsilon_{i k} \delta_{j \ell}+\epsilon_{j \ell} \delta_{i k}+\epsilon_{i \ell} \delta_{j k}+\epsilon_{j k} \delta_{i \ell}\right),
\label{eq:odd}
\end{align}
where $\epsilon_{ij}$ is the 2D Levi-Civita tensor with $\epsilon_{xx}=\epsilon_{yy}=0$ and $\epsilon_{xy}=-\epsilon_{yx}=1$.
The above viscosity tensor $\eta_{{\rm o},ijk\ell}$ is parity-even because both $\sigma_{ij}$ and $v_{k\ell}$ are parity-even, whereas terms that include odd number of $\epsilon_{ij}$ are parity-odd.
Hence, it is concluded from Eq. (\ref{eq:odd}) that $\eta_{\rm o}$ exists only if both time-reversal and parity symmetries are broken~\cite{banerjee2017}.

Using the Poisson-Bracket approach, Markovich \textit{et al.}\ presented a microscopic Hamiltonian theory for odd viscosity and showed that odd viscosity is present both in 2D and 3D systems~\cite{markovich2021}.
The relation between the angular momentum density $\boldsymbol{\ell}$ of rotating particles and odd viscosity was shown to be $\boldsymbol{\ell}=\mathbf{I}\cdot\boldsymbol{\tau}/\Gamma$ in the steady-state, which agrees with the hydrodynamic derivation~\cite{banerjee2017}.
Here, $\mathbf{I}$ is the momenta of inertia tensor, $\boldsymbol{\tau}$ is the torque density, and $\Gamma$ is the rotational friction coefficient of a self-spinning particle.
On the other hand, Khain \textit{et al.}\ systematically studied all possible viscosity coefficients that violate parity in a 3D fluid~\cite{khain2022} and showed that, in some cases, their obtained coefficients correspond to $\boldsymbol{\ell}$ obtained in Ref.~\cite{markovich2021}.

Moreover, the Green-Kubo formula that relates the odd viscosity $\eta_{\rm o}$ to the stress tensor was derived as~\cite{epstein2020,han2021,hargus2020}
\begin{align}
\eta_{\rm o} \sim \int_0^\infty dt\, 
&\left[\langle \sigma_{xx}(t) \sigma_{yx}(0) \rangle
-\langle \sigma_{yx}(t) \sigma_{xx}(0) \rangle \right. \nonumber\\
+&\left.\langle \sigma_{xy}(t) \sigma_{yy}(0) \rangle
-\langle \sigma_{yy}(t) \sigma_{xy}(0) \rangle\right].
\label{eq:oddGK}
\end{align}
When the time-reversal symmetry exists, i.e., $\boldsymbol{\sigma}(t)=\boldsymbol{\sigma}(-t)$, and 
the time translational invariance holds, i.e., $\langle\boldsymbol{\sigma}(t)\boldsymbol{\sigma}(t^\prime)\rangle=\langle\boldsymbol{\sigma}(t-t^\prime)\boldsymbol{\sigma}(0)\rangle$~\cite{doi2013}, the above odd viscosity $\eta_{\rm o}$ vanishes.
This means that the violation of the time-reversal symmetry is essential for the existence of odd viscosity, and the active chiral system with broken symmetries inherently possesses the odd transport coefficient~\cite{banerjee2017,markovich2021}.
By using molecular dynamics simulations and the Green-Kubo formula in Eq.~(\ref{eq:oddGK}), the odd viscosity has been quantitatively measured~\cite{hargus2020,han2021}.

\subsection{Unidirectional edge waves at fluid boundaries}

From the experimental point of view, odd viscosity was measured for a fluid consisting of self-spinning particles~\cite{soni2019,yang2020,yang2021,zhao2021}.
Soni \textit{et al.}\ considered an active chiral fluid that includes spinning colloidal magnets and studied the fluid flow focusing on its surface dynamics~\cite{soni2019}.
They found that unidirectional waves emerged at the fluid boundary and further related the surface tension to odd viscosity, which is the first experimental verification of odd viscosity~\cite{soni2019}.
Similar robust surface flows were also observed at the macroscopic scale, e.g., active chiral granular systems with centimeter-scale toys called Hexbug~\cite{yang2020} or gear-like particles~\cite{yang2021}.
Later, by taking into account the inter-particle hydrodynamic lubrication, the first normal stress difference was related to odd viscosity in the sheared active chiral system~\cite{zhao2021}.

\begin{figure}[tbh]
\centering
\includegraphics[scale=.08]{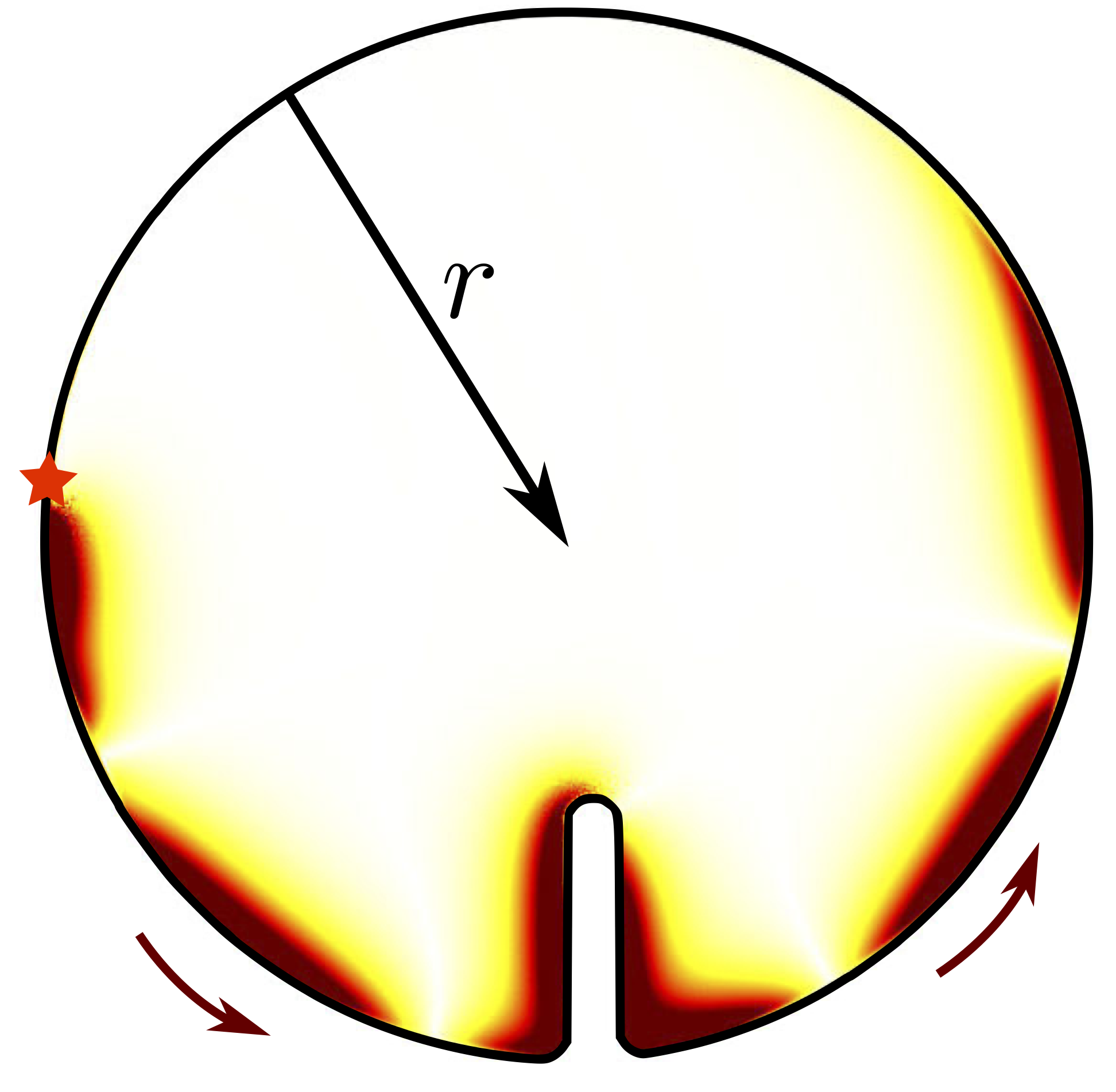}
\caption{
Topological waves in fluids with odd viscosity.
The color shows density deviations.
Reprinted figure with permission from [A.\ Souslov, K.\ Dasbiswas, M.\ Fruchart, S.\ Vaikuntanathan, and V.\ Vitelli, Phys.\ Rev.\ Letts.\ \textbf{122}, 128001 (2019)]~\cite{souslov2019}.
Copyright (2019) by the American Physical Society.
\label{fig:edge}
}
\end{figure}

In quantum systems, odd viscosity has been discussed in relation to topological systems, such as quantum Hall fluids~\cite{avron1995}, whereas in classical systems, the viscosity has rarely been investigated.
Recently, it was shown that the odd viscosity characterizes topological edge modes even in a classical fluid with odd viscosity~\cite{souslov2019,tauber2019,tauber2020}.
As shown in Fig.~\ref{fig:edge},
Souslov \textit{et al.}\ demonstrated that the topological properties of linear waves in a fluid are affected by odd viscosity, and the number of chiral edge states depends on the signs of both odd viscosity and the rotational property of the fluid~\cite{souslov2019}.
They also found that the behavior can be related to a bulk topological invariant Chern number given by
\begin{align}
\mathcal{C} = {\rm sign}(\eta_{\rm o}) + {\rm sign}(\omega),
\end{align}
where $\omega$ is the intrinsic rotation angular frequency of the fluid constituent.
Later, it was shown that edge modes depend on the boundary conditions of the fluids~\cite{tauber2019,tauber2020}.

\subsection{Hydrodynamic effects due to odd viscosity}

Next, we explain the hydrodynamic consequences of odd viscosity, which have been examined in different literatures~\cite{avron1998,ganeshan2017,souslov2020,epstein2020,lapa2014}.
Through the momentum balance equation at low Reynolds number, $\nabla\cdot\boldsymbol{\sigma} -\nabla p = 0$, where inertia is negligible~\cite{doi2013}, one can obtain the hydrodynamic equation for a 2D incompressible fluid with odd viscosity as~\cite{avron1998}
\begin{align}
\eta \nabla^2\mathbf{v}   + \eta_{\rm o} \nabla^2 \boldsymbol{\epsilon}\cdot\mathbf{v} -\nabla p =0,
\label{eq:oddstokes}
\end{align}
where $\eta$ is the 2D shear viscosity and the 2D version of the incompressibility condition $\nabla\cdot\mathbf{v}=0$ of Eq.~(\ref{eq:incompressible}) is assumed.
The second term on the left-hand side of Eq.~(\ref{eq:oddstokes}) is a new contribution due to nonvanishing odd viscosity.
Since the antisymmetric tensor $\boldsymbol{\epsilon}$ accounts for the clockwise rotation by $\pi/2$, odd viscosity contributes to the fluid flow perpendicular to the one generated by the shear viscosity $\eta$.

The hydrodynamic forces acting on various objects have been studied theoretically for a 2D incompressible fluid in the presence of odd viscosity~\cite{ganeshan2017,souslov2020,lapa2014}.
Ganeshan \textit{et al.}\ showed that if boundary conditions depend only on the velocity field, the hydrodynamic force 
does not depend on $\eta_{\rm o}$~\cite{ganeshan2017}.
The force exerted on a unit length of a contour of the object is given by the traction force $f_j=n_i\sigma_{ij}$ where $\mathbf{n}$ is a unit vector normal to the contour in the direction of the fluid.
From the traction due to odd viscosity $f_{{\rm o},j} = 2\eta_{\rm o} \partial_s v_j$ with $\mathbf{s}=-\boldsymbol{\epsilon}\cdot\mathbf{n}$, the force on the object becomes~\cite{ganeshan2017}
\begin{align}
F_{{\rm o},j} = 2\eta_{\rm o}\int ds\, \partial_s v_j=0,
\label{eq:noforce}
\end{align}
which means that the net force acting on an arbitrarily shaped object does not depend on $\eta_{\rm o}$~\cite{ganeshan2017}.
This implies that one should include appropriate boundary conditions in a 2D incompressible fluid to reveal the presence of odd viscosity~\cite{ganeshan2017,souslov2020,lapa2014}.
For example, an expanding bubble with a no-stress boundary condition has been considered, and it was shown that the odd viscosity is responsible for a torque acting on the bubble~\cite{souslov2020,lapa2014,ganeshan2017}.

\begin{figure}[tbh]
\begin{center}
\includegraphics[scale=0.1]{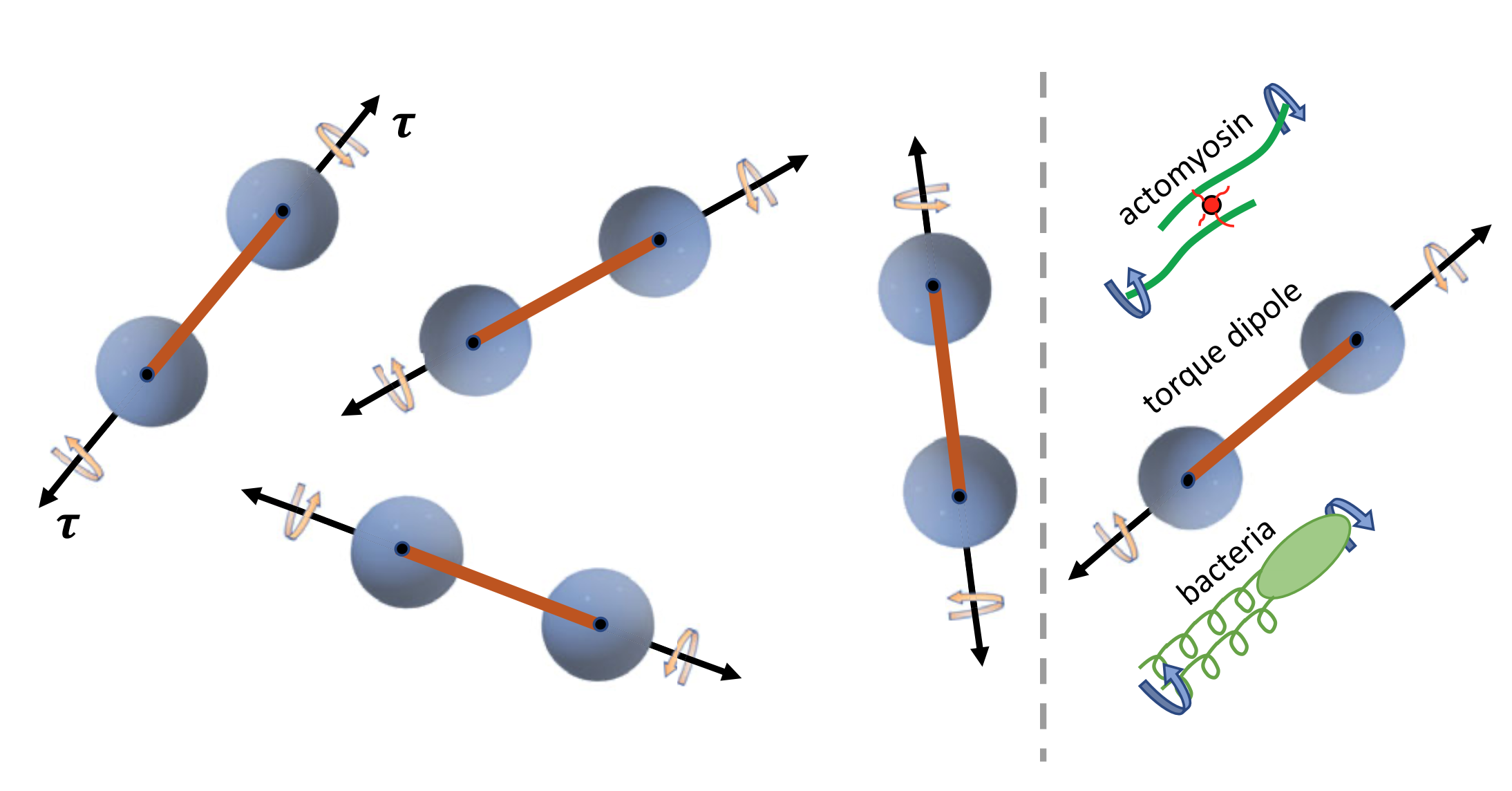}
\end{center}
\caption{
(Left) A fluid of torque dipoles that represents inhomogeneous odd viscosity. 
(Right) Torque dipole as a model for toque exerted by bacteria and for a myosin twisting two actin filaments.
Reprinted figure with permission from [T.\ Markovich and T.\ C.\ Lubensky, Phys.\ Rev.\ Letts.\ \textbf{127}, 048001 (2021)]~\cite{markovich2021}.
Copyright (2021) by the American Physical Society.
\label{fig:torquedipole}
}
\end{figure}

\subsection{Odd viscosity in biological systems}

Active chiral systems are abundant in living systems where the energy-consuming agents and their inherent asymmetry play essential roles.
For instance, biological nanomachines such as ion pumps break both the time-reversal and parity symmetries due to ATP-driven motions and autonomous rotation, which would give rise to odd viscosity in biological membranes~\cite{banerjee2017}. 
Furthermore, in microscopic approaches, it was shown that the odd viscosity also exists in 3D fluids, which extends the applicability of odd viscosity in a living matter such as actomyosin gels 
[see Fig.~\ref{fig:torquedipole}(Right)]~\cite{markovich2021}.

As mentioned in Sec.~\ref{subsec:continuum}, no external force acts on biological nanomachines, and hence the force-free or torque-free conditions should be taken into account for enzymes and micromachines~\cite{mikhailov2015,najafi2004,golestanian2008} or rotary proteins [see Fig.~\ref{fig:torquedipole}(Left)]~\cite{markovich2021}, respectively.
In living systems, moreover, heterogeneity plays an important role, and hence odd viscosity can vary in space~\cite{markovich2021}.
Despite these recent developments in the theory of odd viscosity~\cite{banerjee2017,markovich2021,souslov2019,ganeshan2017,souslov2020,lapa2014,epstein2020}, no experiment observes odd viscosity in living systems.
This is because experimental protocols that allow for the measurement of odd viscosity in a biological context are still sparse, and further theoretical studies are needed to relate odd viscosity to actual physical phenomena.

\begin{figure}[tbh]
\begin{center}
\includegraphics[scale=.2]{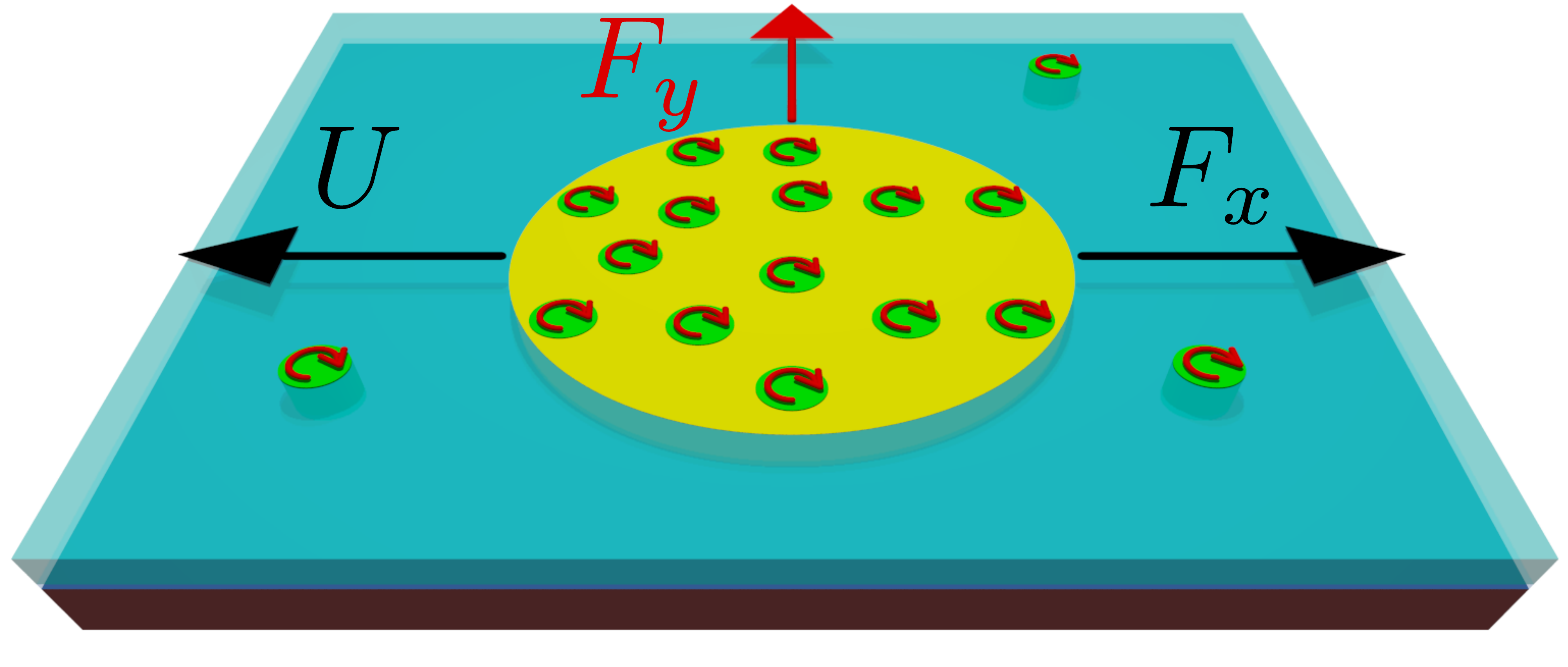}
\end{center}
\caption{
Schematic drawing of a fluid membrane (blue), which is
flat, thin, incompressible, and supported by a rigid substrate (brown).
The membrane has a 2D even (shear) viscosity and odd viscosity. 
The odd viscosity reflects the presence of active
rotor proteins (green) within the membrane that accumulate inside
the liquid domain.
Adapted from [Y.\ Hosaka, S.\ Komura, and D.\ Andelman, Phys.\ Rev.\ E \textbf{104}, 064613 (2021)]~\cite{hosaka2021_2} under CC BY 4.0.
\label{fig:hosaka2021_2}
}
\end{figure}

Hosaka \textit{et al.}\ discussed hydrodynamic forces acting on a 2D liquid domain that moves laterally within a supported fluid membrane in the presence of odd viscosity, as shown in 
Fig.~\ref{fig:hosaka2021_2}~\cite{hosaka2021_2}.
Since active rotating proteins can accumulate inside the domain, they focused on the difference in odd viscosity between the inside and outside of the domain.
By taking into account the momentum leakage from a 2D incompressible fluid to the underlying substrate, they analytically obtained the fluid flow induced by the lateral domain motion and calculated the lift force as well as the drag force acting on the moving liquid domain.
In the presence of an odd viscosity difference, the flow field due to the domain motion is rotated with respect to its direction.
In contrast to the passive case without odd viscosity, the lateral lift arises in the active case when the in and out odd viscosities are different. 
It was shown that the in-out contrast in the odd viscosity leads to nonreciprocal hydrodynamic responses of an active liquid domain.

\begin{figure*}[tbh]
\begin{center}
\includegraphics[scale=.35]{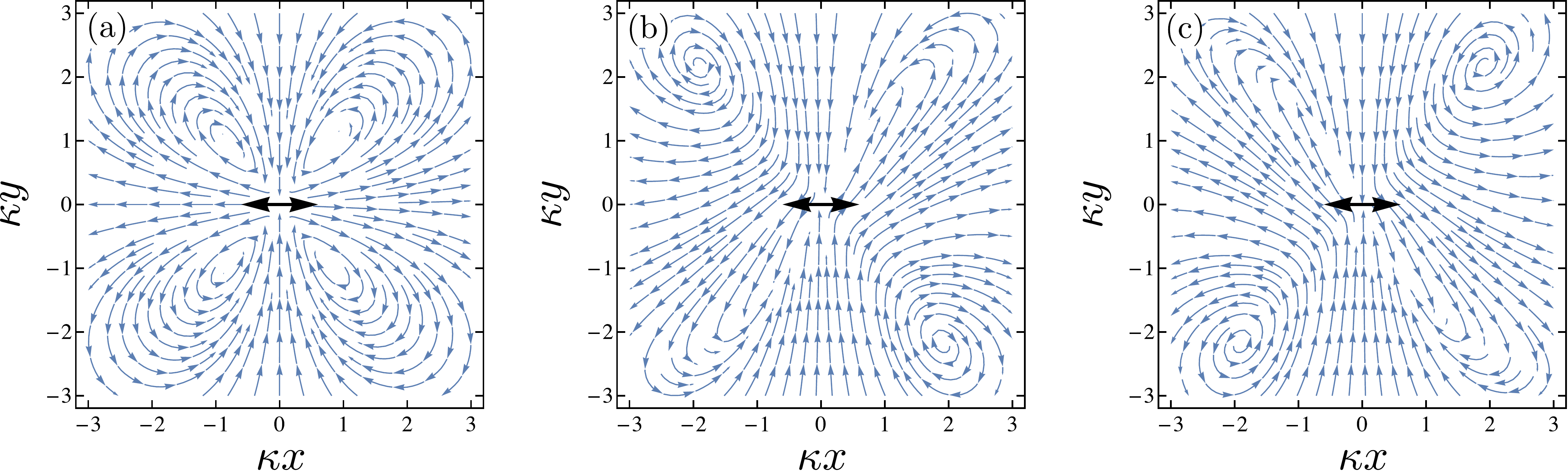}
\end{center}
\caption{
Streamlines of the velocity generated by a hydrodynamic force dipole for (a) $\eta_{\rm o}=0$, (b) $\eta_{\rm o}/\eta=3$, and (c) $\eta_{\rm o}/\eta=-3$, where $\eta_{\rm o}$ and $\eta$ are the 2D odd and shear viscosities, respectively. 
Reprinted figure with permission from [Y.\ Hosaka, S.\ Komura, and D.\ Andelman, Phys.\ Rev.\ E \textbf{13}, 042610 (2021)]~\cite{hosaka2021}.
Copyright (2021) by the American Physical Society.
\label{fig:hosaka2021}
}
\end{figure*}

Moreover, Hosaka \textit{et al.}\ discussed the linear hydrodynamic response of a 2D active chiral compressible (rather than incompressible) fluid with odd viscosity~\cite{hosaka2021}. 
The odd viscosity coefficient represents broken time-reversal and parity symmetries in the 2D fluid and characterizes the deviation of the system from a passive fluid. 
Taking into account the hydrodynamic coupling to the underlying bulk fluid, they obtained the odd viscosity-dependent mobility tensor, which is responsible for the nonreciprocal hydrodynamic response to a point force or a force dipole.
For the passive fluid without odd viscosity, the streamlines have an axial symmetry along the parallel and perpendicular to the force dipole direction, as shown in Fig.~\ref{fig:hosaka2021}(a).
When the odd viscosity is finite, however, flows perpendicular to the applied forces become dominant, and both mirror symmetries are broken, as shown in Figs.~\ref{fig:hosaka2021}(b) and (c).
Furthermore, they considered a finite-size disk moving laterally in the 2D fluid and demonstrated that the disk experiences a nondissipative lift force in addition to the dissipative drag one.

\section{Future Perspectives}
\label{sec:perspectives}

In the future, the active force dipole model can be extended to active rotating proteins that exert torque along their axes rather than force dipoles.
Such a theory can provide a basis to study the parity-breaking effects on the nonequilibrium transport phenomena such as diffusion enhancement and chemotaxis.
It would also be useful to investigate how these physical quantities are modified due to the concentration of biological nanomachines.
A possible approach is to consider stresses that are exerted by biological nanomachines on surrounding fluids so that the effective viscosity of the fluid is modified by the nanomachine concentration (see, e.g., Ref.~\cite{oppenheimer2009}).
Moreover, it would be important to investigate in detail hydrodynamic effects accompanying functional conformational transitions in all-atom or coarse-grained molecular dynamics simulations for specific enzymes and protein machines.

It is possible to consider active force dipoles to model directional cytoplasmic streaming flows observed mainly in plant cells and known to play roles in their growth~\cite{goldstein2015}.
By introducing force dipoles in the vicinity of a cellular membrane and taking into account hydrodynamic flows induced by the dipoles, one can obtain the macroscopic flow, which can be compared with cytoplasmic streaming.
In addition, the flow may exhibit unidirectional transports that would depend on the order parameter defined by the average direction of the dipoles.
These edge currents have been observed in an odd-viscous fluid, and the odd transport coefficient is expected to play an important role in the model.

In Ref.~\cite{hosaka2021}, the obtained drag and lift forces are valid for a small size object as the body force densities in the Lorentz reciprocal theorem were ignored.
To obtain the expressions for a large size object, one has to consider other approaches, e.g., numerical simulations of the boundary integral equations~\cite{pozrikidis1992} or the full derivation of the solutions of the hydrodynamic equations for a 2D chiral fluid~\cite{hayakawa2000}.

Generally, it is useful to reveal the existence of odd viscosity in microscopic approaches.
Numerically, active chiral systems with rotating constituents have been investigated in light of phase separation dynamics~\cite{nguyen2014} and an inertial lift force~\cite{goto2015}.
However, there have been fewer studies that aim at extracting the odd transport coefficient in such systems.
On the other hand, the collective behavior of enzymes in active chiral media deserves attention.
In the presence of thermal noise, one has to consider the generalization of the fluctuation-dissipation theorem as well as the Langevin equation with multiplicative noise when the friction coefficient depends on the position~\cite{lau2007}.
These interesting questions are left for future investigations.

\section*{Acknowledgments}

We thank D.\ Andelman, A.\ S.\ Mikhailov, I.\ Sou, and K.\ Yasuda for useful discussions.
S.K.\ acknowledges the support by the startup fund of Wenzhou Institute, University of Chinese Academy 
of Sciences (No.\ WIUCASQD2021041).

\bibliographystyle{apsrev4-2}
\bibliography{myref}

\begin{thebibliography}{93}%
\makeatletter
\providecommand \@ifxundefined [1]{%
 \@ifx{#1\undefined}
}%
\providecommand \@ifnum [1]{%
 \ifnum #1\expandafter \@firstoftwo
 \else \expandafter \@secondoftwo
 \fi
}%
\providecommand \@ifx [1]{%
 \ifx #1\expandafter \@firstoftwo
 \else \expandafter \@secondoftwo
 \fi
}%
\providecommand \natexlab [1]{#1}%
\providecommand \enquote  [1]{``#1''}%
\providecommand \bibnamefont  [1]{#1}%
\providecommand \bibfnamefont [1]{#1}%
\providecommand \citenamefont [1]{#1}%
\providecommand \href@noop [0]{\@secondoftwo}%
\providecommand \href [0]{\begingroup \@sanitize@url \@href}%
\providecommand \@href[1]{\@@startlink{#1}\@@href}%
\providecommand \@@href[1]{\endgroup#1\@@endlink}%
\providecommand \@sanitize@url [0]{\catcode `\\12\catcode `\$12\catcode
  `\&12\catcode `\#12\catcode `\^12\catcode `\_12\catcode `\%12\relax}%
\providecommand \@@startlink[1]{}%
\providecommand \@@endlink[0]{}%
\providecommand \url  [0]{\begingroup\@sanitize@url \@url }%
\providecommand \@url [1]{\endgroup\@href {#1}{\urlprefix }}%
\providecommand \urlprefix  [0]{URL }%
\providecommand \Eprint [0]{\href }%
\providecommand \doibase [0]{https://doi.org/}%
\providecommand \selectlanguage [0]{\@gobble}%
\providecommand \bibinfo  [0]{\@secondoftwo}%
\providecommand \bibfield  [0]{\@secondoftwo}%
\providecommand \translation [1]{[#1]}%
\providecommand \BibitemOpen [0]{}%
\providecommand \bibitemStop [0]{}%
\providecommand \bibitemNoStop [0]{.\EOS\space}%
\providecommand \EOS [0]{\spacefactor3000\relax}%
\providecommand \BibitemShut  [1]{\csname bibitem#1\endcsname}%
\let\auto@bib@innerbib\@empty
\bibitem [{\citenamefont {Alberts}\ \emph {et~al.}(2008)\citenamefont
  {Alberts}, \citenamefont {Johnson}, \citenamefont {Lewis}, \citenamefont
  {Raff}, \citenamefont {Roberts},\ and\ \citenamefont {Walter}}]{albertsbook}%
  \BibitemOpen
  \bibfield  {author} {\bibinfo {author} {\bibfnamefont {B.}~\bibnamefont
  {Alberts}}, \bibinfo {author} {\bibfnamefont {A.}~\bibnamefont {Johnson}},
  \bibinfo {author} {\bibfnamefont {J.}~\bibnamefont {Lewis}}, \bibinfo
  {author} {\bibfnamefont {M.}~\bibnamefont {Raff}}, \bibinfo {author}
  {\bibfnamefont {K.}~\bibnamefont {Roberts}},\ and\ \bibinfo {author}
  {\bibfnamefont {P.}~\bibnamefont {Walter}},\ }\href@noop {} {\emph {\bibinfo
  {title} {Molecular Biology of the Cell}}}\ (\bibinfo  {publisher} {Garland
  Science},\ \bibinfo {address} {New York},\ \bibinfo {year}
  {2008})\BibitemShut {NoStop}%
\bibitem [{\citenamefont {Phillips}\ \emph {et~al.}(2012)\citenamefont
  {Phillips}, \citenamefont {Kondev}, \citenamefont {Theriot}, \citenamefont
  {Garcia},\ and\ \citenamefont {Orme}}]{phillips2012}%
  \BibitemOpen
  \bibfield  {author} {\bibinfo {author} {\bibfnamefont {R.}~\bibnamefont
  {Phillips}}, \bibinfo {author} {\bibfnamefont {J.}~\bibnamefont {Kondev}},
  \bibinfo {author} {\bibfnamefont {J.}~\bibnamefont {Theriot}}, \bibinfo
  {author} {\bibfnamefont {H.~G.}\ \bibnamefont {Garcia}},\ and\ \bibinfo
  {author} {\bibfnamefont {N.}~\bibnamefont {Orme}},\ }\href@noop {} {\emph
  {\bibinfo {title} {Physical Biology of the Cell}}}\ (\bibinfo  {publisher}
  {Garland Science},\ \bibinfo {address} {New York},\ \bibinfo {year}
  {2012})\BibitemShut {NoStop}%
\bibitem [{\citenamefont {Lu}\ \emph {et~al.}(1998)\citenamefont {Lu},
  \citenamefont {Xun},\ and\ \citenamefont {Xie}}]{lu1998}%
  \BibitemOpen
  \bibfield  {author} {\bibinfo {author} {\bibfnamefont {H.~P.}\ \bibnamefont
  {Lu}}, \bibinfo {author} {\bibfnamefont {L.}~\bibnamefont {Xun}},\ and\
  \bibinfo {author} {\bibfnamefont {X.~S.}\ \bibnamefont {Xie}},\ }\href@noop
  {} {\bibfield  {journal} {\bibinfo  {journal} {Science}\ }\textbf {\bibinfo
  {volume} {282}},\ \bibinfo {pages} {1877} (\bibinfo {year}
  {1998})}\BibitemShut {NoStop}%
\bibitem [{\citenamefont {Kou}\ \emph {et~al.}(2005)\citenamefont {Kou},
  \citenamefont {Cherayil}, \citenamefont {Min}, \citenamefont {English},\ and\
  \citenamefont {Xie}}]{kou2005}%
  \BibitemOpen
  \bibfield  {author} {\bibinfo {author} {\bibfnamefont {S.~C.}\ \bibnamefont
  {Kou}}, \bibinfo {author} {\bibfnamefont {B.~J.}\ \bibnamefont {Cherayil}},
  \bibinfo {author} {\bibfnamefont {W.}~\bibnamefont {Min}}, \bibinfo {author}
  {\bibfnamefont {B.~P.}\ \bibnamefont {English}},\ and\ \bibinfo {author}
  {\bibfnamefont {X.~S.}\ \bibnamefont {Xie}},\ }\href@noop {} {\bibfield
  {journal} {\bibinfo  {journal} {J. Phys. Chem. B}\ }\textbf {\bibinfo
  {volume} {109}},\ \bibinfo {pages} {19068} (\bibinfo {year}
  {2005})}\BibitemShut {NoStop}%
\bibitem [{\citenamefont {English}\ \emph {et~al.}(2006)\citenamefont
  {English}, \citenamefont {Min}, \citenamefont {Van~Oijen}, \citenamefont
  {Lee}, \citenamefont {Luo}, \citenamefont {Sun}, \citenamefont {Cherayil},
  \citenamefont {Kou},\ and\ \citenamefont {Xie}}]{english2006}%
  \BibitemOpen
  \bibfield  {author} {\bibinfo {author} {\bibfnamefont {B.~P.}\ \bibnamefont
  {English}}, \bibinfo {author} {\bibfnamefont {W.}~\bibnamefont {Min}},
  \bibinfo {author} {\bibfnamefont {A.~M.}\ \bibnamefont {Van~Oijen}}, \bibinfo
  {author} {\bibfnamefont {K.~T.}\ \bibnamefont {Lee}}, \bibinfo {author}
  {\bibfnamefont {G.}~\bibnamefont {Luo}}, \bibinfo {author} {\bibfnamefont
  {H.}~\bibnamefont {Sun}}, \bibinfo {author} {\bibfnamefont {B.~J.}\
  \bibnamefont {Cherayil}}, \bibinfo {author} {\bibfnamefont {S.}~\bibnamefont
  {Kou}},\ and\ \bibinfo {author} {\bibfnamefont {X.~S.}\ \bibnamefont {Xie}},\
  }\href@noop {} {\bibfield  {journal} {\bibinfo  {journal} {Nature Chem.
  Bio.}\ }\textbf {\bibinfo {volume} {2}},\ \bibinfo {pages} {87} (\bibinfo
  {year} {2006})}\BibitemShut {NoStop}%
\bibitem [{\citenamefont {Gerstein}\ \emph {et~al.}(1994)\citenamefont
  {Gerstein}, \citenamefont {Lesk},\ and\ \citenamefont
  {Chothia}}]{gerstein1994}%
  \BibitemOpen
  \bibfield  {author} {\bibinfo {author} {\bibfnamefont {M.}~\bibnamefont
  {Gerstein}}, \bibinfo {author} {\bibfnamefont {A.~M.}\ \bibnamefont {Lesk}},\
  and\ \bibinfo {author} {\bibfnamefont {C.}~\bibnamefont {Chothia}},\
  }\href@noop {} {\bibfield  {journal} {\bibinfo  {journal} {Biochemistry}\
  }\textbf {\bibinfo {volume} {33}},\ \bibinfo {pages} {6739} (\bibinfo {year}
  {1994})}\BibitemShut {NoStop}%
\bibitem [{\citenamefont {Or{\"a}dd}\ \emph {et~al.}(2021)\citenamefont
  {Or{\"a}dd}, \citenamefont {Ravishankar}, \citenamefont {Goodman},
  \citenamefont {Rogne}, \citenamefont {Backman}, \citenamefont {Duelli},
  \citenamefont {N.~Pedersen}, \citenamefont {Levantino}, \citenamefont
  {Wulff}, \citenamefont {Wolf-Watz},\ and\ \citenamefont
  {Anderson}}]{oradd2021tracking}%
  \BibitemOpen
  \bibfield  {author} {\bibinfo {author} {\bibfnamefont {F.}~\bibnamefont
  {Or{\"a}dd}}, \bibinfo {author} {\bibfnamefont {H.}~\bibnamefont
  {Ravishankar}}, \bibinfo {author} {\bibfnamefont {J.}~\bibnamefont
  {Goodman}}, \bibinfo {author} {\bibfnamefont {P.}~\bibnamefont {Rogne}},
  \bibinfo {author} {\bibfnamefont {L.}~\bibnamefont {Backman}}, \bibinfo
  {author} {\bibfnamefont {A.}~\bibnamefont {Duelli}}, \bibinfo {author}
  {\bibfnamefont {M.}~\bibnamefont {N.~Pedersen}}, \bibinfo {author}
  {\bibfnamefont {M.}~\bibnamefont {Levantino}}, \bibinfo {author}
  {\bibfnamefont {M.}~\bibnamefont {Wulff}}, \bibinfo {author} {\bibfnamefont
  {M.}~\bibnamefont {Wolf-Watz}},\ and\ \bibinfo {author} {\bibfnamefont
  {M.}~\bibnamefont {Anderson}},\ }\href@noop {} {\bibfield  {journal}
  {\bibinfo  {journal} {Sci. Adv.}\ }\textbf {\bibinfo {volume} {7}},\ \bibinfo
  {pages} {eabi5514} (\bibinfo {year} {2021})}\BibitemShut {NoStop}%
\bibitem [{\citenamefont {Togashi}\ and\ \citenamefont
  {Mikhailov}(2007)}]{togashi2007}%
  \BibitemOpen
  \bibfield  {author} {\bibinfo {author} {\bibfnamefont {Y.}~\bibnamefont
  {Togashi}}\ and\ \bibinfo {author} {\bibfnamefont {A.~S.}\ \bibnamefont
  {Mikhailov}},\ }\href@noop {} {\bibfield  {journal} {\bibinfo  {journal}
  {Proc. Natl. Acad. Sci. (USA)}\ }\textbf {\bibinfo {volume} {104}},\ \bibinfo
  {pages} {8697} (\bibinfo {year} {2007})}\BibitemShut {NoStop}%
\bibitem [{\citenamefont {Echeverria}\ \emph {et~al.}(2011)\citenamefont
  {Echeverria}, \citenamefont {Togashi}, \citenamefont {Mikhailov},\ and\
  \citenamefont {Kapral}}]{echeverria2011}%
  \BibitemOpen
  \bibfield  {author} {\bibinfo {author} {\bibfnamefont {C.}~\bibnamefont
  {Echeverria}}, \bibinfo {author} {\bibfnamefont {Y.}~\bibnamefont {Togashi}},
  \bibinfo {author} {\bibfnamefont {A.~S.}\ \bibnamefont {Mikhailov}},\ and\
  \bibinfo {author} {\bibfnamefont {R.}~\bibnamefont {Kapral}},\ }\href@noop {}
  {\bibfield  {journal} {\bibinfo  {journal} {{P}hys.\ {C}hem.\ {C}hem.\
  {P}hys.}\ }\textbf {\bibinfo {volume} {13}},\ \bibinfo {pages} {10527}
  (\bibinfo {year} {2011})}\BibitemShut {NoStop}%
\bibitem [{\citenamefont {Aviram}\ \emph {et~al.}(2018)\citenamefont {Aviram},
  \citenamefont {Pirchi}, \citenamefont {Mazal}, \citenamefont {Barak},
  \citenamefont {Riven},\ and\ \citenamefont {Haran}}]{aviram2018}%
  \BibitemOpen
  \bibfield  {author} {\bibinfo {author} {\bibfnamefont {H.~Y.}\ \bibnamefont
  {Aviram}}, \bibinfo {author} {\bibfnamefont {M.}~\bibnamefont {Pirchi}},
  \bibinfo {author} {\bibfnamefont {H.}~\bibnamefont {Mazal}}, \bibinfo
  {author} {\bibfnamefont {Y.}~\bibnamefont {Barak}}, \bibinfo {author}
  {\bibfnamefont {I.}~\bibnamefont {Riven}},\ and\ \bibinfo {author}
  {\bibfnamefont {G.}~\bibnamefont {Haran}},\ }\href@noop {} {\bibfield
  {journal} {\bibinfo  {journal} {Proc. Natl. Acad. Sci. (USA)}\ }\textbf
  {\bibinfo {volume} {115}},\ \bibinfo {pages} {3243} (\bibinfo {year}
  {2018})}\BibitemShut {NoStop}%
\bibitem [{\citenamefont {Noji}\ \emph {et~al.}(1997)\citenamefont {Noji},
  \citenamefont {Yasuda}, \citenamefont {Yoshida},\ and\ \citenamefont
  {Kinosita}}]{noji1997}%
  \BibitemOpen
  \bibfield  {author} {\bibinfo {author} {\bibfnamefont {H.}~\bibnamefont
  {Noji}}, \bibinfo {author} {\bibfnamefont {R.}~\bibnamefont {Yasuda}},
  \bibinfo {author} {\bibfnamefont {M.}~\bibnamefont {Yoshida}},\ and\ \bibinfo
  {author} {\bibfnamefont {K.}~\bibnamefont {Kinosita}},\ }\href@noop {}
  {\bibfield  {journal} {\bibinfo  {journal} {Nature}\ }\textbf {\bibinfo
  {volume} {386}},\ \bibinfo {pages} {299} (\bibinfo {year}
  {1997})}\BibitemShut {NoStop}%
\bibitem [{\citenamefont {Muddana}\ \emph {et~al.}(2010)\citenamefont
  {Muddana}, \citenamefont {Sengupta}, \citenamefont {Mallouk}, \citenamefont
  {Sen},\ and\ \citenamefont {Butler}}]{muddana2010}%
  \BibitemOpen
  \bibfield  {author} {\bibinfo {author} {\bibfnamefont {H.~S.}\ \bibnamefont
  {Muddana}}, \bibinfo {author} {\bibfnamefont {S.}~\bibnamefont {Sengupta}},
  \bibinfo {author} {\bibfnamefont {T.~E.}\ \bibnamefont {Mallouk}}, \bibinfo
  {author} {\bibfnamefont {A.}~\bibnamefont {Sen}},\ and\ \bibinfo {author}
  {\bibfnamefont {P.~J.}\ \bibnamefont {Butler}},\ }\href@noop {} {\bibfield
  {journal} {\bibinfo  {journal} {J. Am. Chem. Soc.}\ }\textbf {\bibinfo
  {volume} {132}},\ \bibinfo {pages} {2110} (\bibinfo {year}
  {2010})}\BibitemShut {NoStop}%
\bibitem [{\citenamefont {Sengupta}\ \emph {et~al.}(2013)\citenamefont
  {Sengupta}, \citenamefont {Dey}, \citenamefont {Muddana}, \citenamefont
  {Tabouillot}, \citenamefont {Ibele}, \citenamefont {Butler},\ and\
  \citenamefont {Sen}}]{sengupta2013}%
  \BibitemOpen
  \bibfield  {author} {\bibinfo {author} {\bibfnamefont {S.}~\bibnamefont
  {Sengupta}}, \bibinfo {author} {\bibfnamefont {K.~K.}\ \bibnamefont {Dey}},
  \bibinfo {author} {\bibfnamefont {H.~S.}\ \bibnamefont {Muddana}}, \bibinfo
  {author} {\bibfnamefont {T.}~\bibnamefont {Tabouillot}}, \bibinfo {author}
  {\bibfnamefont {M.~E.}\ \bibnamefont {Ibele}}, \bibinfo {author}
  {\bibfnamefont {P.~J.}\ \bibnamefont {Butler}},\ and\ \bibinfo {author}
  {\bibfnamefont {A.}~\bibnamefont {Sen}},\ }\href@noop {} {\bibfield
  {journal} {\bibinfo  {journal} {J. Am. Chem. Soc.}\ }\textbf {\bibinfo
  {volume} {135}},\ \bibinfo {pages} {1406} (\bibinfo {year}
  {2013})}\BibitemShut {NoStop}%
\bibitem [{\citenamefont {Riedel}\ \emph {et~al.}(2015)\citenamefont {Riedel},
  \citenamefont {Gabizon}, \citenamefont {Wilson}, \citenamefont {Hamadani},
  \citenamefont {Tsekouras}, \citenamefont {Marqusee}, \citenamefont
  {Press\'{e}},\ and\ \citenamefont {Bustamante}}]{riedel2015}%
  \BibitemOpen
  \bibfield  {author} {\bibinfo {author} {\bibfnamefont {C.}~\bibnamefont
  {Riedel}}, \bibinfo {author} {\bibfnamefont {R.}~\bibnamefont {Gabizon}},
  \bibinfo {author} {\bibfnamefont {C.~A.~M.}\ \bibnamefont {Wilson}}, \bibinfo
  {author} {\bibfnamefont {K.}~\bibnamefont {Hamadani}}, \bibinfo {author}
  {\bibfnamefont {K.}~\bibnamefont {Tsekouras}}, \bibinfo {author}
  {\bibfnamefont {S.}~\bibnamefont {Marqusee}}, \bibinfo {author}
  {\bibfnamefont {S.}~\bibnamefont {Press\'{e}}},\ and\ \bibinfo {author}
  {\bibfnamefont {C.}~\bibnamefont {Bustamante}},\ }\href@noop {} {\bibfield
  {journal} {\bibinfo  {journal} {Nature}\ }\textbf {\bibinfo {volume} {517}},\
  \bibinfo {pages} {227} (\bibinfo {year} {2015})}\BibitemShut {NoStop}%
\bibitem [{\citenamefont {Illien}\ \emph
  {et~al.}(2017{\natexlab{a}})\citenamefont {Illien}, \citenamefont {Zhao},
  \citenamefont {Dey}, \citenamefont {Butler}, \citenamefont {Sen},\ and\
  \citenamefont {Golestanian}}]{illien2017_2}%
  \BibitemOpen
  \bibfield  {author} {\bibinfo {author} {\bibfnamefont {P.}~\bibnamefont
  {Illien}}, \bibinfo {author} {\bibfnamefont {X.}~\bibnamefont {Zhao}},
  \bibinfo {author} {\bibfnamefont {K.}~\bibnamefont {Dey}}, \bibinfo {author}
  {\bibfnamefont {P.~J.}\ \bibnamefont {Butler}}, \bibinfo {author}
  {\bibfnamefont {A.}~\bibnamefont {Sen}},\ and\ \bibinfo {author}
  {\bibfnamefont {R.}~\bibnamefont {Golestanian}},\ }\href@noop {} {\bibfield
  {journal} {\bibinfo  {journal} {Nano Lett.}\ }\textbf {\bibinfo {volume}
  {17}},\ \bibinfo {pages} {4415} (\bibinfo {year}
  {2017}{\natexlab{a}})}\BibitemShut {NoStop}%
\bibitem [{\citenamefont {Zhao}\ \emph {et~al.}(2017)\citenamefont {Zhao},
  \citenamefont {Dey}, \citenamefont {Jeganathan}, \citenamefont {Butler},
  \citenamefont {C{\'o}rdova-Figueroa},\ and\ \citenamefont {Sen}}]{zhao2017}%
  \BibitemOpen
  \bibfield  {author} {\bibinfo {author} {\bibfnamefont {X.}~\bibnamefont
  {Zhao}}, \bibinfo {author} {\bibfnamefont {K.~K.}\ \bibnamefont {Dey}},
  \bibinfo {author} {\bibfnamefont {S.}~\bibnamefont {Jeganathan}}, \bibinfo
  {author} {\bibfnamefont {P.~J.}\ \bibnamefont {Butler}}, \bibinfo {author}
  {\bibfnamefont {U.~M.}\ \bibnamefont {C{\'o}rdova-Figueroa}},\ and\ \bibinfo
  {author} {\bibfnamefont {A.}~\bibnamefont {Sen}},\ }\href@noop {} {\bibfield
  {journal} {\bibinfo  {journal} {Nano Lett.}\ }\textbf {\bibinfo {volume}
  {17}},\ \bibinfo {pages} {4807} (\bibinfo {year} {2017})}\BibitemShut
  {NoStop}%
\bibitem [{\citenamefont {Jee}\ \emph {et~al.}(2018)\citenamefont {Jee},
  \citenamefont {Dutta}, \citenamefont {Cho}, \citenamefont {Tlusty},\ and\
  \citenamefont {Granick}}]{jee2018}%
  \BibitemOpen
  \bibfield  {author} {\bibinfo {author} {\bibfnamefont {A.-Y.}\ \bibnamefont
  {Jee}}, \bibinfo {author} {\bibfnamefont {S.}~\bibnamefont {Dutta}}, \bibinfo
  {author} {\bibfnamefont {Y.-K.}\ \bibnamefont {Cho}}, \bibinfo {author}
  {\bibfnamefont {T.}~\bibnamefont {Tlusty}},\ and\ \bibinfo {author}
  {\bibfnamefont {S.}~\bibnamefont {Granick}},\ }\href@noop {} {\bibfield
  {journal} {\bibinfo  {journal} {Proc. Natl. Acad. Sci. (USA)}\ }\textbf
  {\bibinfo {volume} {115}},\ \bibinfo {pages} {14} (\bibinfo {year}
  {2018})}\BibitemShut {NoStop}%
\bibitem [{\citenamefont {Dey}(2019)}]{dey2016}%
  \BibitemOpen
  \bibfield  {author} {\bibinfo {author} {\bibfnamefont {K.~K.}\ \bibnamefont
  {Dey}},\ }\href@noop {} {\bibfield  {journal} {\bibinfo  {journal} {Angew.
  Chem. Int. Ed.}\ }\textbf {\bibinfo {volume} {58}},\ \bibinfo {pages} {2208}
  (\bibinfo {year} {2019})}\BibitemShut {NoStop}%
\bibitem [{\citenamefont {Wang}\ \emph {et~al.}(2020)\citenamefont {Wang},
  \citenamefont {Park}, \citenamefont {Dong}, \citenamefont {Kim},
  \citenamefont {Cho}, \citenamefont {Tlusty},\ and\ \citenamefont
  {Granick}}]{wang2020}%
  \BibitemOpen
  \bibfield  {author} {\bibinfo {author} {\bibfnamefont {H.}~\bibnamefont
  {Wang}}, \bibinfo {author} {\bibfnamefont {M.}~\bibnamefont {Park}}, \bibinfo
  {author} {\bibfnamefont {R.}~\bibnamefont {Dong}}, \bibinfo {author}
  {\bibfnamefont {J.}~\bibnamefont {Kim}}, \bibinfo {author} {\bibfnamefont
  {Y.-K.}\ \bibnamefont {Cho}}, \bibinfo {author} {\bibfnamefont
  {T.}~\bibnamefont {Tlusty}},\ and\ \bibinfo {author} {\bibfnamefont
  {S.}~\bibnamefont {Granick}},\ }\href@noop {} {\bibfield  {journal} {\bibinfo
   {journal} {Science}\ }\textbf {\bibinfo {volume} {369}},\ \bibinfo {pages}
  {537} (\bibinfo {year} {2020})}\BibitemShut {NoStop}%
\bibitem [{\citenamefont {MacDonald}\ \emph {et~al.}(2019)\citenamefont
  {MacDonald}, \citenamefont {Price}, \citenamefont {Astumian},\ and\
  \citenamefont {Beves}}]{macdonald2019}%
  \BibitemOpen
  \bibfield  {author} {\bibinfo {author} {\bibfnamefont {T.~S.}\ \bibnamefont
  {MacDonald}}, \bibinfo {author} {\bibfnamefont {W.~S.}\ \bibnamefont
  {Price}}, \bibinfo {author} {\bibfnamefont {R.~D.}\ \bibnamefont
  {Astumian}},\ and\ \bibinfo {author} {\bibfnamefont {J.~E.}\ \bibnamefont
  {Beves}},\ }\href@noop {} {\bibfield  {journal} {\bibinfo  {journal} {Angew.
  Chem.}\ }\textbf {\bibinfo {volume} {131}},\ \bibinfo {pages} {19040}
  (\bibinfo {year} {2019})}\BibitemShut {NoStop}%
\bibitem [{\citenamefont {Rezaei-Ghaleh}\ \emph {et~al.}(2022)\citenamefont
  {Rezaei-Ghaleh}, \citenamefont {Agudo-Canalejo}, \citenamefont {Griesinger},\
  and\ \citenamefont {Golestanian}}]{rezaei2022}%
  \BibitemOpen
  \bibfield  {author} {\bibinfo {author} {\bibfnamefont {N.}~\bibnamefont
  {Rezaei-Ghaleh}}, \bibinfo {author} {\bibfnamefont {J.}~\bibnamefont
  {Agudo-Canalejo}}, \bibinfo {author} {\bibfnamefont {C.}~\bibnamefont
  {Griesinger}},\ and\ \bibinfo {author} {\bibfnamefont {R.}~\bibnamefont
  {Golestanian}},\ }\href@noop {} {\bibfield  {journal} {\bibinfo  {journal}
  {J. Am. Chem. Soc.}\ }\textbf {\bibinfo {volume} {144}},\ \bibinfo {pages}
  {1380} (\bibinfo {year} {2022})}\BibitemShut {NoStop}%
\bibitem [{\citenamefont {Mikhailov}\ and\ \citenamefont
  {Kapral}(2015)}]{mikhailov2015}%
  \BibitemOpen
  \bibfield  {author} {\bibinfo {author} {\bibfnamefont {A.~S.}\ \bibnamefont
  {Mikhailov}}\ and\ \bibinfo {author} {\bibfnamefont {R.}~\bibnamefont
  {Kapral}},\ }\href@noop {} {\bibfield  {journal} {\bibinfo  {journal} {Proc.
  Natl. Acad. Sci. (USA)}\ }\textbf {\bibinfo {volume} {112}},\ \bibinfo
  {pages} {E3639} (\bibinfo {year} {2015})}\BibitemShut {NoStop}%
\bibitem [{\citenamefont {Kapral}\ and\ \citenamefont
  {Mikhailov}(2016)}]{kapral2016}%
  \BibitemOpen
  \bibfield  {author} {\bibinfo {author} {\bibfnamefont {R.}~\bibnamefont
  {Kapral}}\ and\ \bibinfo {author} {\bibfnamefont {A.~S.}\ \bibnamefont
  {Mikhailov}},\ }\href@noop {} {\bibfield  {journal} {\bibinfo  {journal}
  {Physica D}\ }\textbf {\bibinfo {volume} {318--319}},\ \bibinfo {pages} {104}
  (\bibinfo {year} {2016})}\BibitemShut {NoStop}%
\bibitem [{\citenamefont {Golestanian}(2015)}]{golestanian2015}%
  \BibitemOpen
  \bibfield  {author} {\bibinfo {author} {\bibfnamefont {R.}~\bibnamefont
  {Golestanian}},\ }\href@noop {} {\bibfield  {journal} {\bibinfo  {journal}
  {Phys. Rev. Lett.}\ }\textbf {\bibinfo {volume} {115}},\ \bibinfo {pages}
  {108102} (\bibinfo {year} {2015})}\BibitemShut {NoStop}%
\bibitem [{\citenamefont {Illien}\ \emph
  {et~al.}(2017{\natexlab{b}})\citenamefont {Illien}, \citenamefont
  {Adeleke-Larodo},\ and\ \citenamefont {Golestanian}}]{illien2017}%
  \BibitemOpen
  \bibfield  {author} {\bibinfo {author} {\bibfnamefont {P.}~\bibnamefont
  {Illien}}, \bibinfo {author} {\bibfnamefont {T.}~\bibnamefont
  {Adeleke-Larodo}},\ and\ \bibinfo {author} {\bibfnamefont {R.}~\bibnamefont
  {Golestanian}},\ }\href@noop {} {\bibfield  {journal} {\bibinfo  {journal}
  {EPL}\ }\textbf {\bibinfo {volume} {119}},\ \bibinfo {pages} {40002}
  (\bibinfo {year} {2017}{\natexlab{b}})}\BibitemShut {NoStop}%
\bibitem [{\citenamefont {Xu}\ \emph {et~al.}(2019)\citenamefont {Xu},
  \citenamefont {Ross}, \citenamefont {Valdez},\ and\ \citenamefont
  {Sen}}]{xu2019}%
  \BibitemOpen
  \bibfield  {author} {\bibinfo {author} {\bibfnamefont {M.}~\bibnamefont
  {Xu}}, \bibinfo {author} {\bibfnamefont {J.~L.}\ \bibnamefont {Ross}},
  \bibinfo {author} {\bibfnamefont {L.}~\bibnamefont {Valdez}},\ and\ \bibinfo
  {author} {\bibfnamefont {A.}~\bibnamefont {Sen}},\ }\href@noop {} {\bibfield
  {journal} {\bibinfo  {journal} {Phys. Rev. Lett.}\ }\textbf {\bibinfo
  {volume} {123}},\ \bibinfo {pages} {128101} (\bibinfo {year}
  {2019})}\BibitemShut {NoStop}%
\bibitem [{\citenamefont {Guo}\ \emph {et~al.}(2014)\citenamefont {Guo},
  \citenamefont {Ehrlicher}, \citenamefont {Jensen}, \citenamefont {Renz},
  \citenamefont {Moore}, \citenamefont {Goldman}, \citenamefont
  {Lippincott-Schwartz}, \citenamefont {MacKintosh},\ and\ \citenamefont
  {Weitz}}]{guo2014}%
  \BibitemOpen
  \bibfield  {author} {\bibinfo {author} {\bibfnamefont {M.}~\bibnamefont
  {Guo}}, \bibinfo {author} {\bibfnamefont {A.~J.}\ \bibnamefont {Ehrlicher}},
  \bibinfo {author} {\bibfnamefont {M.~H.}\ \bibnamefont {Jensen}}, \bibinfo
  {author} {\bibfnamefont {M.}~\bibnamefont {Renz}}, \bibinfo {author}
  {\bibfnamefont {J.~R.}\ \bibnamefont {Moore}}, \bibinfo {author}
  {\bibfnamefont {R.~D.}\ \bibnamefont {Goldman}}, \bibinfo {author}
  {\bibfnamefont {J.}~\bibnamefont {Lippincott-Schwartz}}, \bibinfo {author}
  {\bibfnamefont {F.~C.}\ \bibnamefont {MacKintosh}},\ and\ \bibinfo {author}
  {\bibfnamefont {D.~A.}\ \bibnamefont {Weitz}},\ }\href@noop {} {\bibfield
  {journal} {\bibinfo  {journal} {Cell}\ }\textbf {\bibinfo {volume} {158}},\
  \bibinfo {pages} {822} (\bibinfo {year} {2014})}\BibitemShut {NoStop}%
\bibitem [{\citenamefont {Parry}\ \emph {et~al.}(2014)\citenamefont {Parry},
  \citenamefont {Surovtsev}, \citenamefont {Cabeen}, \citenamefont {O'Hem},
  \citenamefont {Dufresne},\ and\ \citenamefont {Jacobs-Wagner}}]{parry2014}%
  \BibitemOpen
  \bibfield  {author} {\bibinfo {author} {\bibfnamefont {B.~R.}\ \bibnamefont
  {Parry}}, \bibinfo {author} {\bibfnamefont {I.~V.}\ \bibnamefont
  {Surovtsev}}, \bibinfo {author} {\bibfnamefont {M.~T.}\ \bibnamefont
  {Cabeen}}, \bibinfo {author} {\bibfnamefont {C.~S.}\ \bibnamefont {O'Hem}},
  \bibinfo {author} {\bibfnamefont {E.~R.}\ \bibnamefont {Dufresne}},\ and\
  \bibinfo {author} {\bibfnamefont {C.}~\bibnamefont {Jacobs-Wagner}},\
  }\href@noop {} {\bibfield  {journal} {\bibinfo  {journal} {Cell}\ }\textbf
  {\bibinfo {volume} {156}},\ \bibinfo {pages} {183} (\bibinfo {year}
  {2014})}\BibitemShut {NoStop}%
\bibitem [{\citenamefont {Saintillan}(2018)}]{saintillan2018}%
  \BibitemOpen
  \bibfield  {author} {\bibinfo {author} {\bibfnamefont {D.}~\bibnamefont
  {Saintillan}},\ }\href@noop {} {\bibfield  {journal} {\bibinfo  {journal}
  {Annu. Rev. Fluid Mech.}\ }\textbf {\bibinfo {volume} {50}},\ \bibinfo
  {pages} {563} (\bibinfo {year} {2018})}\BibitemShut {NoStop}%
\bibitem [{\citenamefont {Landau}\ and\ \citenamefont
  {Lifshitz}(1987)}]{Landau1987}%
  \BibitemOpen
  \bibfield  {author} {\bibinfo {author} {\bibfnamefont {L.~D.}\ \bibnamefont
  {Landau}}\ and\ \bibinfo {author} {\bibfnamefont {E.~M.}\ \bibnamefont
  {Lifshitz}},\ }\href@noop {} {\emph {\bibinfo {title} {Fluid Mechanics}}}\
  (\bibinfo  {publisher} {Pergamon Press},\ \bibinfo {address} {Oxford},\
  \bibinfo {year} {1987})\BibitemShut {NoStop}%
\bibitem [{\citenamefont {Nishizawa}\ \emph {et~al.}(2017)\citenamefont
  {Nishizawa}, \citenamefont {Fujiwara}, \citenamefont {Ikenaga}, \citenamefont
  {Nakajo}, \citenamefont {Yanagisawa},\ and\ \citenamefont
  {Mizuno}}]{nishizawa2017}%
  \BibitemOpen
  \bibfield  {author} {\bibinfo {author} {\bibfnamefont {K.}~\bibnamefont
  {Nishizawa}}, \bibinfo {author} {\bibfnamefont {K.}~\bibnamefont {Fujiwara}},
  \bibinfo {author} {\bibfnamefont {M.}~\bibnamefont {Ikenaga}}, \bibinfo
  {author} {\bibfnamefont {N.}~\bibnamefont {Nakajo}}, \bibinfo {author}
  {\bibfnamefont {M.}~\bibnamefont {Yanagisawa}},\ and\ \bibinfo {author}
  {\bibfnamefont {D.}~\bibnamefont {Mizuno}},\ }\href@noop {} {\bibfield
  {journal} {\bibinfo  {journal} {Sci. Rep.}\ }\textbf {\bibinfo {volume}
  {7}},\ \bibinfo {pages} {1} (\bibinfo {year} {2017})}\BibitemShut {NoStop}%
\bibitem [{\citenamefont {L{\'o}pez}\ \emph {et~al.}(2015)\citenamefont
  {L{\'o}pez}, \citenamefont {Gachelin}, \citenamefont {Douarche},
  \citenamefont {Auradou},\ and\ \citenamefont {Cl{\'e}ment}}]{lopez2015}%
  \BibitemOpen
  \bibfield  {author} {\bibinfo {author} {\bibfnamefont {H.~M.}\ \bibnamefont
  {L{\'o}pez}}, \bibinfo {author} {\bibfnamefont {J.}~\bibnamefont {Gachelin}},
  \bibinfo {author} {\bibfnamefont {C.}~\bibnamefont {Douarche}}, \bibinfo
  {author} {\bibfnamefont {H.}~\bibnamefont {Auradou}},\ and\ \bibinfo {author}
  {\bibfnamefont {E.}~\bibnamefont {Cl{\'e}ment}},\ }\href@noop {} {\bibfield
  {journal} {\bibinfo  {journal} {Phys. Rev. Lett.}\ }\textbf {\bibinfo
  {volume} {115}},\ \bibinfo {pages} {028301} (\bibinfo {year}
  {2015})}\BibitemShut {NoStop}%
\bibitem [{\citenamefont {Rafa{\"\i}}\ \emph {et~al.}(2010)\citenamefont
  {Rafa{\"\i}}, \citenamefont {Jibuti},\ and\ \citenamefont
  {Peyla}}]{rafai2010}%
  \BibitemOpen
  \bibfield  {author} {\bibinfo {author} {\bibfnamefont {S.}~\bibnamefont
  {Rafa{\"\i}}}, \bibinfo {author} {\bibfnamefont {L.}~\bibnamefont {Jibuti}},\
  and\ \bibinfo {author} {\bibfnamefont {P.}~\bibnamefont {Peyla}},\
  }\href@noop {} {\bibfield  {journal} {\bibinfo  {journal} {Phys. Rev. Lett.}\
  }\textbf {\bibinfo {volume} {104}},\ \bibinfo {pages} {098102} (\bibinfo
  {year} {2010})}\BibitemShut {NoStop}%
\bibitem [{\citenamefont {Sumino}\ \emph {et~al.}(2012)\citenamefont {Sumino},
  \citenamefont {Nagai}, \citenamefont {Shitaka}, \citenamefont {Tanaka},
  \citenamefont {Yoshikawa}, \citenamefont {Chat{\'e}},\ and\ \citenamefont
  {Oiwa}}]{sumino2012}%
  \BibitemOpen
  \bibfield  {author} {\bibinfo {author} {\bibfnamefont {Y.}~\bibnamefont
  {Sumino}}, \bibinfo {author} {\bibfnamefont {K.~H.}\ \bibnamefont {Nagai}},
  \bibinfo {author} {\bibfnamefont {Y.}~\bibnamefont {Shitaka}}, \bibinfo
  {author} {\bibfnamefont {D.}~\bibnamefont {Tanaka}}, \bibinfo {author}
  {\bibfnamefont {K.}~\bibnamefont {Yoshikawa}}, \bibinfo {author}
  {\bibfnamefont {H.}~\bibnamefont {Chat{\'e}}},\ and\ \bibinfo {author}
  {\bibfnamefont {K.}~\bibnamefont {Oiwa}},\ }\href@noop {} {\bibfield
  {journal} {\bibinfo  {journal} {Nature}\ }\textbf {\bibinfo {volume} {483}},\
  \bibinfo {pages} {448} (\bibinfo {year} {2012})}\BibitemShut {NoStop}%
\bibitem [{\citenamefont {Tabe}\ and\ \citenamefont
  {Yokoyama}(2003)}]{tabe2003}%
  \BibitemOpen
  \bibfield  {author} {\bibinfo {author} {\bibfnamefont {Y.}~\bibnamefont
  {Tabe}}\ and\ \bibinfo {author} {\bibfnamefont {H.}~\bibnamefont
  {Yokoyama}},\ }\href@noop {} {\bibfield  {journal} {\bibinfo  {journal} {Nat.
  Mater.}\ }\textbf {\bibinfo {volume} {2}},\ \bibinfo {pages} {806} (\bibinfo
  {year} {2003})}\BibitemShut {NoStop}%
\bibitem [{\citenamefont {Yamauchi}\ \emph {et~al.}(2020)\citenamefont
  {Yamauchi}, \citenamefont {Hayata}, \citenamefont {Uwamichi}, \citenamefont
  {Ozawa},\ and\ \citenamefont {Kawaguchi}}]{yamauchi2020}%
  \BibitemOpen
  \bibfield  {author} {\bibinfo {author} {\bibfnamefont {L.}~\bibnamefont
  {Yamauchi}}, \bibinfo {author} {\bibfnamefont {T.}~\bibnamefont {Hayata}},
  \bibinfo {author} {\bibfnamefont {M.}~\bibnamefont {Uwamichi}}, \bibinfo
  {author} {\bibfnamefont {T.}~\bibnamefont {Ozawa}},\ and\ \bibinfo {author}
  {\bibfnamefont {K.}~\bibnamefont {Kawaguchi}},\ }\href@noop {} {\bibfield
  {journal} {\bibinfo  {journal} {arXiv preprint arXiv:2008.10852}\ } (\bibinfo
  {year} {2020})}\BibitemShut {NoStop}%
\bibitem [{\citenamefont {Beppu}\ \emph {et~al.}(2021)\citenamefont {Beppu},
  \citenamefont {Izri}, \citenamefont {Sato}, \citenamefont {Yamanishi},
  \citenamefont {Sumino},\ and\ \citenamefont {Maeda}}]{beppu2021}%
  \BibitemOpen
  \bibfield  {author} {\bibinfo {author} {\bibfnamefont {K.}~\bibnamefont
  {Beppu}}, \bibinfo {author} {\bibfnamefont {Z.}~\bibnamefont {Izri}},
  \bibinfo {author} {\bibfnamefont {T.}~\bibnamefont {Sato}}, \bibinfo {author}
  {\bibfnamefont {Y.}~\bibnamefont {Yamanishi}}, \bibinfo {author}
  {\bibfnamefont {Y.}~\bibnamefont {Sumino}},\ and\ \bibinfo {author}
  {\bibfnamefont {Y.~T.}\ \bibnamefont {Maeda}},\ }\href@noop {} {\bibfield
  {journal} {\bibinfo  {journal} {Proc. Natl. Acad. Sci. (USA)}\ }\textbf
  {\bibinfo {volume} {118}} (\bibinfo {year} {2021})}\BibitemShut {NoStop}%
\bibitem [{\citenamefont {Banerjee}\ \emph {et~al.}(2017)\citenamefont
  {Banerjee}, \citenamefont {Souslov}, \citenamefont {Abanov},\ and\
  \citenamefont {Vitelli}}]{banerjee2017}%
  \BibitemOpen
  \bibfield  {author} {\bibinfo {author} {\bibfnamefont {D.}~\bibnamefont
  {Banerjee}}, \bibinfo {author} {\bibfnamefont {A.}~\bibnamefont {Souslov}},
  \bibinfo {author} {\bibfnamefont {A.~G.}\ \bibnamefont {Abanov}},\ and\
  \bibinfo {author} {\bibfnamefont {V.}~\bibnamefont {Vitelli}},\ }\href@noop
  {} {\bibfield  {journal} {\bibinfo  {journal} {Nat. Commun.}\ }\textbf
  {\bibinfo {volume} {8}},\ \bibinfo {pages} {1573} (\bibinfo {year}
  {2017})}\BibitemShut {NoStop}%
\bibitem [{\citenamefont {Gompper}\ \emph {et~al.}(2020)\citenamefont
  {Gompper}, \citenamefont {Winkler}, \citenamefont {Speck}, \citenamefont
  {Solon}, \citenamefont {Nardini}, \citenamefont {Peruani}, \citenamefont
  {L{\"o}wen}, \citenamefont {Golestanian}, \citenamefont {Kaupp},
  \citenamefont {Alvarez} \emph {et~al.}}]{gompper2020}%
  \BibitemOpen
  \bibfield  {author} {\bibinfo {author} {\bibfnamefont {G.}~\bibnamefont
  {Gompper}}, \bibinfo {author} {\bibfnamefont {R.~G.}\ \bibnamefont
  {Winkler}}, \bibinfo {author} {\bibfnamefont {T.}~\bibnamefont {Speck}},
  \bibinfo {author} {\bibfnamefont {A.}~\bibnamefont {Solon}}, \bibinfo
  {author} {\bibfnamefont {C.}~\bibnamefont {Nardini}}, \bibinfo {author}
  {\bibfnamefont {F.}~\bibnamefont {Peruani}}, \bibinfo {author} {\bibfnamefont
  {H.}~\bibnamefont {L{\"o}wen}}, \bibinfo {author} {\bibfnamefont
  {R.}~\bibnamefont {Golestanian}}, \bibinfo {author} {\bibfnamefont {U.~B.}\
  \bibnamefont {Kaupp}}, \bibinfo {author} {\bibfnamefont {L.}~\bibnamefont
  {Alvarez}}, \emph {et~al.},\ }\href@noop {} {\bibfield  {journal} {\bibinfo
  {journal} {J. Phys.: Condens. Matter}\ }\textbf {\bibinfo {volume} {32}},\
  \bibinfo {pages} {193001} (\bibinfo {year} {2020})}\BibitemShut {NoStop}%
\bibitem [{\citenamefont {Doi}\ and\ \citenamefont {Edwards}(1986)}]{doi1988}%
  \BibitemOpen
  \bibfield  {author} {\bibinfo {author} {\bibfnamefont {M.}~\bibnamefont
  {Doi}}\ and\ \bibinfo {author} {\bibfnamefont {S.~F.}\ \bibnamefont
  {Edwards}},\ }\href@noop {} {\emph {\bibinfo {title} {The Theory of Polymer
  Dynamics}}}\ (\bibinfo  {publisher} {Oxford University Press, New York},\
  \bibinfo {year} {1986})\BibitemShut {NoStop}%
\bibitem [{\citenamefont {Kim}\ and\ \citenamefont {Karrila}(2013)}]{kim2013}%
  \BibitemOpen
  \bibfield  {author} {\bibinfo {author} {\bibfnamefont {S.}~\bibnamefont
  {Kim}}\ and\ \bibinfo {author} {\bibfnamefont {S.~J.}\ \bibnamefont
  {Karrila}},\ }\href@noop {} {\emph {\bibinfo {title} {Microhydrodynamics:
  Principles and Selected Applications}}}\ (\bibinfo  {publisher} {Courier
  Corporation},\ \bibinfo {year} {2013})\BibitemShut {NoStop}%
\bibitem [{\citenamefont {Oppenheimer}\ \emph {et~al.}(2019)\citenamefont
  {Oppenheimer}, \citenamefont {Stein},\ and\ \citenamefont
  {Shelley}}]{oppenheimer2019}%
  \BibitemOpen
  \bibfield  {author} {\bibinfo {author} {\bibfnamefont {N.}~\bibnamefont
  {Oppenheimer}}, \bibinfo {author} {\bibfnamefont {D.~B.}\ \bibnamefont
  {Stein}},\ and\ \bibinfo {author} {\bibfnamefont {M.~J.}\ \bibnamefont
  {Shelley}},\ }\href@noop {} {\bibfield  {journal} {\bibinfo  {journal} {Phys.
  Rev. Lett.}\ }\textbf {\bibinfo {volume} {123}},\ \bibinfo {pages} {148101}
  (\bibinfo {year} {2019})}\BibitemShut {NoStop}%
\bibitem [{\citenamefont {Lenz}\ \emph {et~al.}(2003)\citenamefont {Lenz},
  \citenamefont {Joanny}, \citenamefont {J{\"u}licher},\ and\ \citenamefont
  {Prost}}]{lenz2003}%
  \BibitemOpen
  \bibfield  {author} {\bibinfo {author} {\bibfnamefont {P.}~\bibnamefont
  {Lenz}}, \bibinfo {author} {\bibfnamefont {J.-F.}\ \bibnamefont {Joanny}},
  \bibinfo {author} {\bibfnamefont {F.}~\bibnamefont {J{\"u}licher}},\ and\
  \bibinfo {author} {\bibfnamefont {J.}~\bibnamefont {Prost}},\ }\href@noop {}
  {\bibfield  {journal} {\bibinfo  {journal} {Phys. Rev. Lett.}\ }\textbf
  {\bibinfo {volume} {91}},\ \bibinfo {pages} {108104} (\bibinfo {year}
  {2003})}\BibitemShut {NoStop}%
\bibitem [{\citenamefont {Saffman}\ and\ \citenamefont
  {Delbr{\"u}ck}(1975)}]{saffman1975}%
  \BibitemOpen
  \bibfield  {author} {\bibinfo {author} {\bibfnamefont {P.~G.}\ \bibnamefont
  {Saffman}}\ and\ \bibinfo {author} {\bibfnamefont {M.}~\bibnamefont
  {Delbr{\"u}ck}},\ }\href@noop {} {\bibfield  {journal} {\bibinfo  {journal}
  {Proc. Natl. Acad. Sci. (USA)}\ }\textbf {\bibinfo {volume} {72}},\ \bibinfo
  {pages} {3111} (\bibinfo {year} {1975})}\BibitemShut {NoStop}%
\bibitem [{\citenamefont {Saffman}(1976)}]{saffman1976}%
  \BibitemOpen
  \bibfield  {author} {\bibinfo {author} {\bibfnamefont {P.~G.}\ \bibnamefont
  {Saffman}},\ }\href@noop {} {\bibfield  {journal} {\bibinfo  {journal} {J.
  Fluid Mech.}\ }\textbf {\bibinfo {volume} {73}},\ \bibinfo {pages} {593}
  (\bibinfo {year} {1976})}\BibitemShut {NoStop}%
\bibitem [{\citenamefont {Evans}\ and\ \citenamefont
  {Sackmann}(1988)}]{evans1988}%
  \BibitemOpen
  \bibfield  {author} {\bibinfo {author} {\bibfnamefont {E.}~\bibnamefont
  {Evans}}\ and\ \bibinfo {author} {\bibfnamefont {E.}~\bibnamefont
  {Sackmann}},\ }\href@noop {} {\bibfield  {journal} {\bibinfo  {journal} {J.
  Fluid Mech.}\ }\textbf {\bibinfo {volume} {194}},\ \bibinfo {pages} {553}
  (\bibinfo {year} {1988})}\BibitemShut {NoStop}%
\bibitem [{\citenamefont {Henle}\ and\ \citenamefont
  {Levine}(2010)}]{henle2010}%
  \BibitemOpen
  \bibfield  {author} {\bibinfo {author} {\bibfnamefont {M.~L.}\ \bibnamefont
  {Henle}}\ and\ \bibinfo {author} {\bibfnamefont {A.~J.}\ \bibnamefont
  {Levine}},\ }\href@noop {} {\bibfield  {journal} {\bibinfo  {journal} {Phys.
  Rev. E}\ }\textbf {\bibinfo {volume} {81}},\ \bibinfo {pages} {011905}
  (\bibinfo {year} {2010})}\BibitemShut {NoStop}%
\bibitem [{\citenamefont {Oppenheimer}\ and\ \citenamefont
  {Diamant}(2009)}]{oppenheimer2009}%
  \BibitemOpen
  \bibfield  {author} {\bibinfo {author} {\bibfnamefont {N.}~\bibnamefont
  {Oppenheimer}}\ and\ \bibinfo {author} {\bibfnamefont {H.}~\bibnamefont
  {Diamant}},\ }\href@noop {} {\bibfield  {journal} {\bibinfo  {journal}
  {Biophys. J.}\ }\textbf {\bibinfo {volume} {96}},\ \bibinfo {pages} {3041}
  (\bibinfo {year} {2009})}\BibitemShut {NoStop}%
\bibitem [{\citenamefont {Oppenheimer}\ and\ \citenamefont
  {Diamant}(2010)}]{oppenheimer2010}%
  \BibitemOpen
  \bibfield  {author} {\bibinfo {author} {\bibfnamefont {N.}~\bibnamefont
  {Oppenheimer}}\ and\ \bibinfo {author} {\bibfnamefont {H.}~\bibnamefont
  {Diamant}},\ }\href@noop {} {\bibfield  {journal} {\bibinfo  {journal} {Phys.
  Rev. E}\ }\textbf {\bibinfo {volume} {82}},\ \bibinfo {pages} {041912}
  (\bibinfo {year} {2010})}\BibitemShut {NoStop}%
\bibitem [{\citenamefont {Ramachandran}\ \emph {et~al.}(2011)\citenamefont
  {Ramachandran}, \citenamefont {Komura}, \citenamefont {Seki},\ and\
  \citenamefont {Gompper}}]{ramachandran2011}%
  \BibitemOpen
  \bibfield  {author} {\bibinfo {author} {\bibfnamefont {S.}~\bibnamefont
  {Ramachandran}}, \bibinfo {author} {\bibfnamefont {S.}~\bibnamefont
  {Komura}}, \bibinfo {author} {\bibfnamefont {K.}~\bibnamefont {Seki}},\ and\
  \bibinfo {author} {\bibfnamefont {G.}~\bibnamefont {Gompper}},\ }\href@noop
  {} {\bibfield  {journal} {\bibinfo  {journal} {Eur. Phys. J. E}\ }\textbf
  {\bibinfo {volume} {34}},\ \bibinfo {pages} {46} (\bibinfo {year}
  {2011})}\BibitemShut {NoStop}%
\bibitem [{\citenamefont {Hosaka}\ \emph
  {et~al.}(2020{\natexlab{a}})\citenamefont {Hosaka}, \citenamefont {Komura},\
  and\ \citenamefont {Mikhailov}}]{hosaka2020_2}%
  \BibitemOpen
  \bibfield  {author} {\bibinfo {author} {\bibfnamefont {Y.}~\bibnamefont
  {Hosaka}}, \bibinfo {author} {\bibfnamefont {S.}~\bibnamefont {Komura}},\
  and\ \bibinfo {author} {\bibfnamefont {A.~S.}\ \bibnamefont {Mikhailov}},\
  }\href@noop {} {\bibfield  {journal} {\bibinfo  {journal} {Soft Matter}\
  }\textbf {\bibinfo {volume} {16}},\ \bibinfo {pages} {10734} (\bibinfo {year}
  {2020}{\natexlab{a}})}\BibitemShut {NoStop}%
\bibitem [{\citenamefont {Koyano}\ \emph {et~al.}(2016)\citenamefont {Koyano},
  \citenamefont {Kitahata},\ and\ \citenamefont {Mikhailov}}]{koyano2016}%
  \BibitemOpen
  \bibfield  {author} {\bibinfo {author} {\bibfnamefont {Y.}~\bibnamefont
  {Koyano}}, \bibinfo {author} {\bibfnamefont {H.}~\bibnamefont {Kitahata}},\
  and\ \bibinfo {author} {\bibfnamefont {A.~S.}\ \bibnamefont {Mikhailov}},\
  }\href@noop {} {\bibfield  {journal} {\bibinfo  {journal} {Phys. Rev. E}\
  }\textbf {\bibinfo {volume} {94}},\ \bibinfo {pages} {022416} (\bibinfo
  {year} {2016})}\BibitemShut {NoStop}%
\bibitem [{\citenamefont {Hosaka}\ \emph {et~al.}(2017)\citenamefont {Hosaka},
  \citenamefont {Yasuda}, \citenamefont {Okamoto},\ and\ \citenamefont
  {Komura}}]{hosaka2017}%
  \BibitemOpen
  \bibfield  {author} {\bibinfo {author} {\bibfnamefont {Y.}~\bibnamefont
  {Hosaka}}, \bibinfo {author} {\bibfnamefont {K.}~\bibnamefont {Yasuda}},
  \bibinfo {author} {\bibfnamefont {R.}~\bibnamefont {Okamoto}},\ and\ \bibinfo
  {author} {\bibfnamefont {S.}~\bibnamefont {Komura}},\ }\href@noop {}
  {\bibfield  {journal} {\bibinfo  {journal} {Phys. Rev. E}\ }\textbf {\bibinfo
  {volume} {95}},\ \bibinfo {pages} {052407} (\bibinfo {year}
  {2017})}\BibitemShut {NoStop}%
\bibitem [{\citenamefont {Manikantan}(2020)}]{manikantan2020}%
  \BibitemOpen
  \bibfield  {author} {\bibinfo {author} {\bibfnamefont {H.}~\bibnamefont
  {Manikantan}},\ }\href@noop {} {\bibfield  {journal} {\bibinfo  {journal}
  {Phys. Rev. Lett.}\ }\textbf {\bibinfo {volume} {125}},\ \bibinfo {pages}
  {268101} (\bibinfo {year} {2020})}\BibitemShut {NoStop}%
\bibitem [{\citenamefont {Bagaria}\ and\ \citenamefont
  {Samanta}(2021)}]{bagaria2021}%
  \BibitemOpen
  \bibfield  {author} {\bibinfo {author} {\bibfnamefont {S.}~\bibnamefont
  {Bagaria}}\ and\ \bibinfo {author} {\bibfnamefont {R.}~\bibnamefont
  {Samanta}},\ }\href@noop {} {\bibfield  {journal} {\bibinfo  {journal} {arXiv
  preprint arXiv:2110.05460}\ } (\bibinfo {year} {2021})}\BibitemShut {NoStop}%
\bibitem [{\citenamefont {Hosaka}\ \emph
  {et~al.}(2020{\natexlab{b}})\citenamefont {Hosaka}, \citenamefont {Komura},\
  and\ \citenamefont {Andelman}}]{hosaka2020}%
  \BibitemOpen
  \bibfield  {author} {\bibinfo {author} {\bibfnamefont {Y.}~\bibnamefont
  {Hosaka}}, \bibinfo {author} {\bibfnamefont {S.}~\bibnamefont {Komura}},\
  and\ \bibinfo {author} {\bibfnamefont {D.}~\bibnamefont {Andelman}},\
  }\href@noop {} {\bibfield  {journal} {\bibinfo  {journal} {Phys. Rev. E}\
  }\textbf {\bibinfo {volume} {101}},\ \bibinfo {pages} {012610} (\bibinfo
  {year} {2020}{\natexlab{b}})}\BibitemShut {NoStop}%
\bibitem [{\citenamefont {Adeleke-Larodo}\ \emph
  {et~al.}(2019{\natexlab{a}})\citenamefont {Adeleke-Larodo}, \citenamefont
  {Illien},\ and\ \citenamefont {Golestanian}}]{adeleke2019}%
  \BibitemOpen
  \bibfield  {author} {\bibinfo {author} {\bibfnamefont {T.}~\bibnamefont
  {Adeleke-Larodo}}, \bibinfo {author} {\bibfnamefont {P.}~\bibnamefont
  {Illien}},\ and\ \bibinfo {author} {\bibfnamefont {R.}~\bibnamefont
  {Golestanian}},\ }\href@noop {} {\bibfield  {journal} {\bibinfo  {journal}
  {Eur. Phys. J. E}\ }\textbf {\bibinfo {volume} {42}},\ \bibinfo {pages} {1}
  (\bibinfo {year} {2019}{\natexlab{a}})}\BibitemShut {NoStop}%
\bibitem [{\citenamefont {Golestanian}(2019)}]{golestanian2019}%
  \BibitemOpen
  \bibfield  {author} {\bibinfo {author} {\bibfnamefont {R.}~\bibnamefont
  {Golestanian}},\ }\href@noop {} {\bibfield  {journal} {\bibinfo  {journal}
  {arXiv preprint arXiv:1909.03747}\ } (\bibinfo {year} {2019})}\BibitemShut
  {NoStop}%
\bibitem [{\citenamefont {Adeleke-Larodo}\ \emph
  {et~al.}(2019{\natexlab{b}})\citenamefont {Adeleke-Larodo}, \citenamefont
  {Agudo-Canalejo},\ and\ \citenamefont {Golestanian}}]{adeleke2019_2}%
  \BibitemOpen
  \bibfield  {author} {\bibinfo {author} {\bibfnamefont {T.}~\bibnamefont
  {Adeleke-Larodo}}, \bibinfo {author} {\bibfnamefont {J.}~\bibnamefont
  {Agudo-Canalejo}},\ and\ \bibinfo {author} {\bibfnamefont {R.}~\bibnamefont
  {Golestanian}},\ }\href@noop {} {\bibfield  {journal} {\bibinfo  {journal}
  {J. Chem. Phys.}\ }\textbf {\bibinfo {volume} {150}},\ \bibinfo {pages}
  {115102} (\bibinfo {year} {2019}{\natexlab{b}})}\BibitemShut {NoStop}%
\bibitem [{\citenamefont {Agudo-Canalejo}\ and\ \citenamefont
  {Golestanian}(2019)}]{agudo2019}%
  \BibitemOpen
  \bibfield  {author} {\bibinfo {author} {\bibfnamefont {J.}~\bibnamefont
  {Agudo-Canalejo}}\ and\ \bibinfo {author} {\bibfnamefont {R.}~\bibnamefont
  {Golestanian}},\ }\href@noop {} {\bibfield  {journal} {\bibinfo  {journal}
  {Phys. Rev. Lett.}\ }\textbf {\bibinfo {volume} {123}},\ \bibinfo {pages}
  {018101} (\bibinfo {year} {2019})}\BibitemShut {NoStop}%
\bibitem [{\citenamefont {Masoud}\ and\ \citenamefont
  {Stone}(2019)}]{masoud2019}%
  \BibitemOpen
  \bibfield  {author} {\bibinfo {author} {\bibfnamefont {H.}~\bibnamefont
  {Masoud}}\ and\ \bibinfo {author} {\bibfnamefont {H.~A.}\ \bibnamefont
  {Stone}},\ }\href@noop {} {\bibfield  {journal} {\bibinfo  {journal} {J.
  Fluid Mech.}\ }\textbf {\bibinfo {volume} {879}},\ \bibinfo {pages} {1}
  (\bibinfo {year} {2019})}\BibitemShut {NoStop}%
\bibitem [{\citenamefont {Ouazan-Reboul}\ \emph {et~al.}(2021)\citenamefont
  {Ouazan-Reboul}, \citenamefont {Agudo-Canalejo},\ and\ \citenamefont
  {Golestanian}}]{ouazan2021}%
  \BibitemOpen
  \bibfield  {author} {\bibinfo {author} {\bibfnamefont {V.}~\bibnamefont
  {Ouazan-Reboul}}, \bibinfo {author} {\bibfnamefont {J.}~\bibnamefont
  {Agudo-Canalejo}},\ and\ \bibinfo {author} {\bibfnamefont {R.}~\bibnamefont
  {Golestanian}},\ }\href@noop {} {\bibfield  {journal} {\bibinfo  {journal}
  {Eur. Phys. J. E}\ }\textbf {\bibinfo {volume} {44}},\ \bibinfo {pages} {113}
  (\bibinfo {year} {2021})}\BibitemShut {NoStop}%
\bibitem [{\citenamefont {Saha}\ \emph {et~al.}(2020)\citenamefont {Saha},
  \citenamefont {Agudo-Canalejo},\ and\ \citenamefont
  {Golestanian}}]{saha2020}%
  \BibitemOpen
  \bibfield  {author} {\bibinfo {author} {\bibfnamefont {S.}~\bibnamefont
  {Saha}}, \bibinfo {author} {\bibfnamefont {J.}~\bibnamefont
  {Agudo-Canalejo}},\ and\ \bibinfo {author} {\bibfnamefont {R.}~\bibnamefont
  {Golestanian}},\ }\href@noop {} {\bibfield  {journal} {\bibinfo  {journal}
  {Phys. Rev. X}\ }\textbf {\bibinfo {volume} {10}},\ \bibinfo {pages} {041009}
  (\bibinfo {year} {2020})}\BibitemShut {NoStop}%
\bibitem [{\citenamefont {Lifshitz}\ and\ \citenamefont
  {Pitaevskii}(1981)}]{lifshitz1981}%
  \BibitemOpen
  \bibfield  {author} {\bibinfo {author} {\bibfnamefont {E.}~\bibnamefont
  {Lifshitz}}\ and\ \bibinfo {author} {\bibfnamefont {L.}~\bibnamefont
  {Pitaevskii}},\ }\href@noop {} {\emph {\bibinfo {title} {Physical
  Kinetics}}}\ (\bibinfo  {publisher} {Pergamon Press},\ \bibinfo {address}
  {Oxford},\ \bibinfo {year} {1981})\BibitemShut {NoStop}%
\bibitem [{\citenamefont {Avron}\ \emph {et~al.}(1995)\citenamefont {Avron},
  \citenamefont {Seiler},\ and\ \citenamefont {Zograf}}]{avron1995}%
  \BibitemOpen
  \bibfield  {author} {\bibinfo {author} {\bibfnamefont {J.}~\bibnamefont
  {Avron}}, \bibinfo {author} {\bibfnamefont {R.}~\bibnamefont {Seiler}},\ and\
  \bibinfo {author} {\bibfnamefont {P.~G.}\ \bibnamefont {Zograf}},\
  }\href@noop {} {\bibfield  {journal} {\bibinfo  {journal} {Phys. Rev. Lett.}\
  }\textbf {\bibinfo {volume} {75}},\ \bibinfo {pages} {697} (\bibinfo {year}
  {1995})}\BibitemShut {NoStop}%
\bibitem [{\citenamefont {Read}(2009)}]{read2009}%
  \BibitemOpen
  \bibfield  {author} {\bibinfo {author} {\bibfnamefont {N.}~\bibnamefont
  {Read}},\ }\href@noop {} {\bibfield  {journal} {\bibinfo  {journal} {Phys.
  Rev. B}\ }\textbf {\bibinfo {volume} {79}},\ \bibinfo {pages} {045308}
  (\bibinfo {year} {2009})}\BibitemShut {NoStop}%
\bibitem [{\citenamefont {Avron}(1998)}]{avron1998}%
  \BibitemOpen
  \bibfield  {author} {\bibinfo {author} {\bibfnamefont {J.~E.}\ \bibnamefont
  {Avron}},\ }\href@noop {} {\bibfield  {journal} {\bibinfo  {journal} {J.
  Stat. Phys.}\ }\textbf {\bibinfo {volume} {92}},\ \bibinfo {pages} {543}
  (\bibinfo {year} {1998})}\BibitemShut {NoStop}%
\bibitem [{\citenamefont {Ganeshan}\ and\ \citenamefont
  {Abanov}(2017)}]{ganeshan2017}%
  \BibitemOpen
  \bibfield  {author} {\bibinfo {author} {\bibfnamefont {S.}~\bibnamefont
  {Ganeshan}}\ and\ \bibinfo {author} {\bibfnamefont {A.~G.}\ \bibnamefont
  {Abanov}},\ }\href@noop {} {\bibfield  {journal} {\bibinfo  {journal} {Phys.
  Rev. Fluids}\ }\textbf {\bibinfo {volume} {2}},\ \bibinfo {pages} {094101}
  (\bibinfo {year} {2017})}\BibitemShut {NoStop}%
\bibitem [{\citenamefont {Epstein}\ and\ \citenamefont
  {Mandadapu}(2020)}]{epstein2020}%
  \BibitemOpen
  \bibfield  {author} {\bibinfo {author} {\bibfnamefont {J.~M.}\ \bibnamefont
  {Epstein}}\ and\ \bibinfo {author} {\bibfnamefont {K.~K.}\ \bibnamefont
  {Mandadapu}},\ }\href@noop {} {\bibfield  {journal} {\bibinfo  {journal}
  {Phys. Rev. E}\ }\textbf {\bibinfo {volume} {101}},\ \bibinfo {pages}
  {052614} (\bibinfo {year} {2020})}\BibitemShut {NoStop}%
\bibitem [{\citenamefont {Markovich}\ and\ \citenamefont
  {Lubensky}(2021)}]{markovich2021}%
  \BibitemOpen
  \bibfield  {author} {\bibinfo {author} {\bibfnamefont {T.}~\bibnamefont
  {Markovich}}\ and\ \bibinfo {author} {\bibfnamefont {T.~C.}\ \bibnamefont
  {Lubensky}},\ }\href@noop {} {\bibfield  {journal} {\bibinfo  {journal}
  {Phys. Rev. Lett.}\ }\textbf {\bibinfo {volume} {127}},\ \bibinfo {pages}
  {048001} (\bibinfo {year} {2021})}\BibitemShut {NoStop}%
\bibitem [{\citenamefont {Khain}\ \emph {et~al.}(2022)\citenamefont {Khain},
  \citenamefont {Scheibner}, \citenamefont {Fruchart},\ and\ \citenamefont
  {Vitelli}}]{khain2022}%
  \BibitemOpen
  \bibfield  {author} {\bibinfo {author} {\bibfnamefont {T.}~\bibnamefont
  {Khain}}, \bibinfo {author} {\bibfnamefont {C.}~\bibnamefont {Scheibner}},
  \bibinfo {author} {\bibfnamefont {M.}~\bibnamefont {Fruchart}},\ and\
  \bibinfo {author} {\bibfnamefont {V.}~\bibnamefont {Vitelli}},\ }\href@noop
  {} {\bibfield  {journal} {\bibinfo  {journal} {J. Fluid Mech.}\ }\textbf
  {\bibinfo {volume} {934}} (\bibinfo {year} {2022})}\BibitemShut {NoStop}%
\bibitem [{\citenamefont {Han}\ \emph {et~al.}(2021)\citenamefont {Han},
  \citenamefont {Fruchart}, \citenamefont {Scheibner}, \citenamefont
  {Vaikuntanathan}, \citenamefont {de~Pablo},\ and\ \citenamefont
  {Vitelli}}]{han2021}%
  \BibitemOpen
  \bibfield  {author} {\bibinfo {author} {\bibfnamefont {M.}~\bibnamefont
  {Han}}, \bibinfo {author} {\bibfnamefont {M.}~\bibnamefont {Fruchart}},
  \bibinfo {author} {\bibfnamefont {C.}~\bibnamefont {Scheibner}}, \bibinfo
  {author} {\bibfnamefont {S.}~\bibnamefont {Vaikuntanathan}}, \bibinfo
  {author} {\bibfnamefont {J.~J.}\ \bibnamefont {de~Pablo}},\ and\ \bibinfo
  {author} {\bibfnamefont {V.}~\bibnamefont {Vitelli}},\ }\href@noop {}
  {\bibfield  {journal} {\bibinfo  {journal} {Nat. Phys.}\ }\textbf {\bibinfo
  {volume} {17}},\ \bibinfo {pages} {1260} (\bibinfo {year}
  {2021})}\BibitemShut {NoStop}%
\bibitem [{\citenamefont {Hargus}\ \emph {et~al.}(2020)\citenamefont {Hargus},
  \citenamefont {Klymko}, \citenamefont {Epstein},\ and\ \citenamefont
  {Mandadapu}}]{hargus2020}%
  \BibitemOpen
  \bibfield  {author} {\bibinfo {author} {\bibfnamefont {C.}~\bibnamefont
  {Hargus}}, \bibinfo {author} {\bibfnamefont {K.}~\bibnamefont {Klymko}},
  \bibinfo {author} {\bibfnamefont {J.~M.}\ \bibnamefont {Epstein}},\ and\
  \bibinfo {author} {\bibfnamefont {K.~K.}\ \bibnamefont {Mandadapu}},\
  }\href@noop {} {\bibfield  {journal} {\bibinfo  {journal} {J. Chem. Phys.}\
  }\textbf {\bibinfo {volume} {152}},\ \bibinfo {pages} {201102} (\bibinfo
  {year} {2020})}\BibitemShut {NoStop}%
\bibitem [{\citenamefont {Doi}(2013)}]{doi2013}%
  \BibitemOpen
  \bibfield  {author} {\bibinfo {author} {\bibfnamefont {M.}~\bibnamefont
  {Doi}},\ }\href@noop {} {\emph {\bibinfo {title} {Soft Matter Physics}}}\
  (\bibinfo  {publisher} {Oxford University Press},\ \bibinfo {year}
  {2013})\BibitemShut {NoStop}%
\bibitem [{\citenamefont {Soni}\ \emph {et~al.}(2019)\citenamefont {Soni},
  \citenamefont {Bililign}, \citenamefont {Magkiriadou}, \citenamefont
  {Sacanna}, \citenamefont {Bartolo}, \citenamefont {Shelley},\ and\
  \citenamefont {Irvine}}]{soni2019}%
  \BibitemOpen
  \bibfield  {author} {\bibinfo {author} {\bibfnamefont {V.}~\bibnamefont
  {Soni}}, \bibinfo {author} {\bibfnamefont {E.~S.}\ \bibnamefont {Bililign}},
  \bibinfo {author} {\bibfnamefont {S.}~\bibnamefont {Magkiriadou}}, \bibinfo
  {author} {\bibfnamefont {S.}~\bibnamefont {Sacanna}}, \bibinfo {author}
  {\bibfnamefont {D.}~\bibnamefont {Bartolo}}, \bibinfo {author} {\bibfnamefont
  {M.~J.}\ \bibnamefont {Shelley}},\ and\ \bibinfo {author} {\bibfnamefont
  {W.~T.~M.}\ \bibnamefont {Irvine}},\ }\href@noop {} {\bibfield  {journal}
  {\bibinfo  {journal} {Nat. Phys.}\ }\textbf {\bibinfo {volume} {15}},\
  \bibinfo {pages} {1188} (\bibinfo {year} {2019})}\BibitemShut {NoStop}%
\bibitem [{\citenamefont {Yang}\ \emph {et~al.}(2020)\citenamefont {Yang},
  \citenamefont {Ren}, \citenamefont {Cheng},\ and\ \citenamefont
  {Zhang}}]{yang2020}%
  \BibitemOpen
  \bibfield  {author} {\bibinfo {author} {\bibfnamefont {X.}~\bibnamefont
  {Yang}}, \bibinfo {author} {\bibfnamefont {C.}~\bibnamefont {Ren}}, \bibinfo
  {author} {\bibfnamefont {K.}~\bibnamefont {Cheng}},\ and\ \bibinfo {author}
  {\bibfnamefont {H.}~\bibnamefont {Zhang}},\ }\href@noop {} {\bibfield
  {journal} {\bibinfo  {journal} {Phys. Rev. E}\ }\textbf {\bibinfo {volume}
  {101}},\ \bibinfo {pages} {022603} (\bibinfo {year} {2020})}\BibitemShut
  {NoStop}%
\bibitem [{\citenamefont {Yang}\ \emph {et~al.}(2021)\citenamefont {Yang},
  \citenamefont {Zhu}, \citenamefont {Liu}, \citenamefont {Liu}, \citenamefont
  {Shi}, \citenamefont {Chen}, \citenamefont {Zheng}, \citenamefont {Ye},\ and\
  \citenamefont {Yang}}]{yang2021}%
  \BibitemOpen
  \bibfield  {author} {\bibinfo {author} {\bibfnamefont {Q.}~\bibnamefont
  {Yang}}, \bibinfo {author} {\bibfnamefont {H.}~\bibnamefont {Zhu}}, \bibinfo
  {author} {\bibfnamefont {P.}~\bibnamefont {Liu}}, \bibinfo {author}
  {\bibfnamefont {R.}~\bibnamefont {Liu}}, \bibinfo {author} {\bibfnamefont
  {Q.}~\bibnamefont {Shi}}, \bibinfo {author} {\bibfnamefont {K.}~\bibnamefont
  {Chen}}, \bibinfo {author} {\bibfnamefont {N.}~\bibnamefont {Zheng}},
  \bibinfo {author} {\bibfnamefont {F.}~\bibnamefont {Ye}},\ and\ \bibinfo
  {author} {\bibfnamefont {M.}~\bibnamefont {Yang}},\ }\href@noop {} {\bibfield
   {journal} {\bibinfo  {journal} {Phys. Rev. Lett.}\ }\textbf {\bibinfo
  {volume} {126}},\ \bibinfo {pages} {198001} (\bibinfo {year}
  {2021})}\BibitemShut {NoStop}%
\bibitem [{\citenamefont {Zhao}\ \emph {et~al.}(2021)\citenamefont {Zhao},
  \citenamefont {Wang}, \citenamefont {Komura}, \citenamefont {Yang},
  \citenamefont {Ye},\ and\ \citenamefont {Seto}}]{zhao2021}%
  \BibitemOpen
  \bibfield  {author} {\bibinfo {author} {\bibfnamefont {Z.}~\bibnamefont
  {Zhao}}, \bibinfo {author} {\bibfnamefont {B.}~\bibnamefont {Wang}}, \bibinfo
  {author} {\bibfnamefont {S.}~\bibnamefont {Komura}}, \bibinfo {author}
  {\bibfnamefont {M.}~\bibnamefont {Yang}}, \bibinfo {author} {\bibfnamefont
  {F.}~\bibnamefont {Ye}},\ and\ \bibinfo {author} {\bibfnamefont
  {R.}~\bibnamefont {Seto}},\ }\href@noop {} {\bibfield  {journal} {\bibinfo
  {journal} {Phys. Rev. Res.}\ }\textbf {\bibinfo {volume} {3}},\ \bibinfo
  {pages} {043229} (\bibinfo {year} {2021})}\BibitemShut {NoStop}%
\bibitem [{\citenamefont {Souslov}\ \emph {et~al.}(2019)\citenamefont
  {Souslov}, \citenamefont {Dasbiswas}, \citenamefont {Fruchart}, \citenamefont
  {Vaikuntanathan},\ and\ \citenamefont {Vitelli}}]{souslov2019}%
  \BibitemOpen
  \bibfield  {author} {\bibinfo {author} {\bibfnamefont {A.}~\bibnamefont
  {Souslov}}, \bibinfo {author} {\bibfnamefont {K.}~\bibnamefont {Dasbiswas}},
  \bibinfo {author} {\bibfnamefont {M.}~\bibnamefont {Fruchart}}, \bibinfo
  {author} {\bibfnamefont {S.}~\bibnamefont {Vaikuntanathan}},\ and\ \bibinfo
  {author} {\bibfnamefont {V.}~\bibnamefont {Vitelli}},\ }\href@noop {}
  {\bibfield  {journal} {\bibinfo  {journal} {Phys. Rev. Lett.}\ }\textbf
  {\bibinfo {volume} {122}},\ \bibinfo {pages} {128001} (\bibinfo {year}
  {2019})}\BibitemShut {NoStop}%
\bibitem [{\citenamefont {Tauber}\ \emph {et~al.}(2019)\citenamefont {Tauber},
  \citenamefont {Delplace},\ and\ \citenamefont {Venaille}}]{tauber2019}%
  \BibitemOpen
  \bibfield  {author} {\bibinfo {author} {\bibfnamefont {C.}~\bibnamefont
  {Tauber}}, \bibinfo {author} {\bibfnamefont {P.}~\bibnamefont {Delplace}},\
  and\ \bibinfo {author} {\bibfnamefont {A.}~\bibnamefont {Venaille}},\
  }\href@noop {} {\bibfield  {journal} {\bibinfo  {journal} {J. Fluid Mech.}\
  }\textbf {\bibinfo {volume} {868}},\ \bibinfo {pages} {R2} (\bibinfo {year}
  {2019})}\BibitemShut {NoStop}%
\bibitem [{\citenamefont {Tauber}\ \emph {et~al.}(2020)\citenamefont {Tauber},
  \citenamefont {Delplace},\ and\ \citenamefont {Venaille}}]{tauber2020}%
  \BibitemOpen
  \bibfield  {author} {\bibinfo {author} {\bibfnamefont {C.}~\bibnamefont
  {Tauber}}, \bibinfo {author} {\bibfnamefont {P.}~\bibnamefont {Delplace}},\
  and\ \bibinfo {author} {\bibfnamefont {A.}~\bibnamefont {Venaille}},\
  }\href@noop {} {\bibfield  {journal} {\bibinfo  {journal} {Phys. Rev. Res.}\
  }\textbf {\bibinfo {volume} {2}},\ \bibinfo {pages} {013147} (\bibinfo {year}
  {2020})}\BibitemShut {NoStop}%
\bibitem [{\citenamefont {Souslov}\ \emph {et~al.}(2020)\citenamefont
  {Souslov}, \citenamefont {Gromov},\ and\ \citenamefont
  {Vitelli}}]{souslov2020}%
  \BibitemOpen
  \bibfield  {author} {\bibinfo {author} {\bibfnamefont {A.}~\bibnamefont
  {Souslov}}, \bibinfo {author} {\bibfnamefont {A.}~\bibnamefont {Gromov}},\
  and\ \bibinfo {author} {\bibfnamefont {V.}~\bibnamefont {Vitelli}},\
  }\href@noop {} {\bibfield  {journal} {\bibinfo  {journal} {Phys. Rev. E}\
  }\textbf {\bibinfo {volume} {101}},\ \bibinfo {pages} {052606} (\bibinfo
  {year} {2020})}\BibitemShut {NoStop}%
\bibitem [{\citenamefont {Lapa}\ and\ \citenamefont {Hughes}(2014)}]{lapa2014}%
  \BibitemOpen
  \bibfield  {author} {\bibinfo {author} {\bibfnamefont {M.~F.}\ \bibnamefont
  {Lapa}}\ and\ \bibinfo {author} {\bibfnamefont {T.~L.}\ \bibnamefont
  {Hughes}},\ }\href@noop {} {\bibfield  {journal} {\bibinfo  {journal} {Phys.
  Rev. E}\ }\textbf {\bibinfo {volume} {89}},\ \bibinfo {pages} {043019}
  (\bibinfo {year} {2014})}\BibitemShut {NoStop}%
\bibitem [{\citenamefont {Najafi}\ and\ \citenamefont
  {Golestanian}(2004)}]{najafi2004}%
  \BibitemOpen
  \bibfield  {author} {\bibinfo {author} {\bibfnamefont {A.}~\bibnamefont
  {Najafi}}\ and\ \bibinfo {author} {\bibfnamefont {R.}~\bibnamefont
  {Golestanian}},\ }\href@noop {} {\bibfield  {journal} {\bibinfo  {journal}
  {Phys. Rev. E}\ }\textbf {\bibinfo {volume} {69}},\ \bibinfo {pages} {062901}
  (\bibinfo {year} {2004})}\BibitemShut {NoStop}%
\bibitem [{\citenamefont {Golestanian}\ and\ \citenamefont
  {Ajdari}(2008)}]{golestanian2008}%
  \BibitemOpen
  \bibfield  {author} {\bibinfo {author} {\bibfnamefont {R.}~\bibnamefont
  {Golestanian}}\ and\ \bibinfo {author} {\bibfnamefont {A.}~\bibnamefont
  {Ajdari}},\ }\href@noop {} {\bibfield  {journal} {\bibinfo  {journal} {Phys.
  Rev. E}\ }\textbf {\bibinfo {volume} {77}},\ \bibinfo {pages} {036308}
  (\bibinfo {year} {2008})}\BibitemShut {NoStop}%
\bibitem [{\citenamefont {Hosaka}\ \emph
  {et~al.}(2021{\natexlab{a}})\citenamefont {Hosaka}, \citenamefont {Komura},\
  and\ \citenamefont {Andelman}}]{hosaka2021_2}%
  \BibitemOpen
  \bibfield  {author} {\bibinfo {author} {\bibfnamefont {Y.}~\bibnamefont
  {Hosaka}}, \bibinfo {author} {\bibfnamefont {S.}~\bibnamefont {Komura}},\
  and\ \bibinfo {author} {\bibfnamefont {D.}~\bibnamefont {Andelman}},\
  }\href@noop {} {\bibfield  {journal} {\bibinfo  {journal} {Phys. Rev. E}\
  }\textbf {\bibinfo {volume} {104}},\ \bibinfo {pages} {064613} (\bibinfo
  {year} {2021}{\natexlab{a}})}\BibitemShut {NoStop}%
\bibitem [{\citenamefont {Hosaka}\ \emph
  {et~al.}(2021{\natexlab{b}})\citenamefont {Hosaka}, \citenamefont {Komura},\
  and\ \citenamefont {Andelman}}]{hosaka2021}%
  \BibitemOpen
  \bibfield  {author} {\bibinfo {author} {\bibfnamefont {Y.}~\bibnamefont
  {Hosaka}}, \bibinfo {author} {\bibfnamefont {S.}~\bibnamefont {Komura}},\
  and\ \bibinfo {author} {\bibfnamefont {D.}~\bibnamefont {Andelman}},\
  }\href@noop {} {\bibfield  {journal} {\bibinfo  {journal} {Phys. Rev. E}\
  }\textbf {\bibinfo {volume} {103}},\ \bibinfo {pages} {042610} (\bibinfo
  {year} {2021}{\natexlab{b}})}\BibitemShut {NoStop}%
\bibitem [{\citenamefont {Goldstein}\ and\ \citenamefont {van~de
  Meent}(2015)}]{goldstein2015}%
  \BibitemOpen
  \bibfield  {author} {\bibinfo {author} {\bibfnamefont {R.~E.}\ \bibnamefont
  {Goldstein}}\ and\ \bibinfo {author} {\bibfnamefont {J.-W.}\ \bibnamefont
  {van~de Meent}},\ }\href@noop {} {\bibfield  {journal} {\bibinfo  {journal}
  {Interface Focus}\ }\textbf {\bibinfo {volume} {5}},\ \bibinfo {pages}
  {20150030} (\bibinfo {year} {2015})}\BibitemShut {NoStop}%
\bibitem [{\citenamefont {Pozrikidis}(1992)}]{pozrikidis1992}%
  \BibitemOpen
  \bibfield  {author} {\bibinfo {author} {\bibfnamefont {C.}~\bibnamefont
  {Pozrikidis}},\ }\href@noop {} {\emph {\bibinfo {title} {Boundary Integral
  and Singularity Methods for Linearized Viscous Flow}}}\ (\bibinfo
  {publisher} {Cambridge University Press},\ \bibinfo {address} {New York},\
  \bibinfo {year} {1992})\BibitemShut {NoStop}%
\bibitem [{\citenamefont {Hayakawa}(2000)}]{hayakawa2000}%
  \BibitemOpen
  \bibfield  {author} {\bibinfo {author} {\bibfnamefont {H.}~\bibnamefont
  {Hayakawa}},\ }\href@noop {} {\bibfield  {journal} {\bibinfo  {journal}
  {Phys. Rev. E}\ }\textbf {\bibinfo {volume} {61}},\ \bibinfo {pages} {5477}
  (\bibinfo {year} {2000})}\BibitemShut {NoStop}%
\bibitem [{\citenamefont {Nguyen}\ \emph {et~al.}(2014)\citenamefont {Nguyen},
  \citenamefont {Klotsa}, \citenamefont {Engel},\ and\ \citenamefont
  {Glotzer}}]{nguyen2014}%
  \BibitemOpen
  \bibfield  {author} {\bibinfo {author} {\bibfnamefont {N.~H.}\ \bibnamefont
  {Nguyen}}, \bibinfo {author} {\bibfnamefont {D.}~\bibnamefont {Klotsa}},
  \bibinfo {author} {\bibfnamefont {M.}~\bibnamefont {Engel}},\ and\ \bibinfo
  {author} {\bibfnamefont {S.~C.}\ \bibnamefont {Glotzer}},\ }\href@noop {}
  {\bibfield  {journal} {\bibinfo  {journal} {Phys. Rev. Lett.}\ }\textbf
  {\bibinfo {volume} {112}},\ \bibinfo {pages} {075701} (\bibinfo {year}
  {2014})}\BibitemShut {NoStop}%
\bibitem [{\citenamefont {Goto}\ and\ \citenamefont {Tanaka}(2015)}]{goto2015}%
  \BibitemOpen
  \bibfield  {author} {\bibinfo {author} {\bibfnamefont {Y.}~\bibnamefont
  {Goto}}\ and\ \bibinfo {author} {\bibfnamefont {H.}~\bibnamefont {Tanaka}},\
  }\href@noop {} {\bibfield  {journal} {\bibinfo  {journal} {Nat. Commun.}\
  }\textbf {\bibinfo {volume} {6}},\ \bibinfo {pages} {1} (\bibinfo {year}
  {2015})}\BibitemShut {NoStop}%
\bibitem [{\citenamefont {Lau}\ and\ \citenamefont {Lubensky}(2007)}]{lau2007}%
  \BibitemOpen
  \bibfield  {author} {\bibinfo {author} {\bibfnamefont {A.~W.}\ \bibnamefont
  {Lau}}\ and\ \bibinfo {author} {\bibfnamefont {T.~C.}\ \bibnamefont
  {Lubensky}},\ }\href@noop {} {\bibfield  {journal} {\bibinfo  {journal}
  {Phys. Rev. E}\ }\textbf {\bibinfo {volume} {76}},\ \bibinfo {pages} {011123}
  (\bibinfo {year} {2007})}\BibitemShut {NoStop}%
\end{thebibliography}%

\end{document}